\documentclass[twocolumn,prb,superscriptaddress,amssymb]{revtex4-2}
\usepackage{graphicx}
\usepackage{amsmath}
\usepackage{xcolor}
\graphicspath{{figures/}}
\def\bk{{\mathbf{k}}}
\def\bq{{\mathbf{q}}}
\def\sgn{{\textnormal{sgn}}}
\def\im{{\textnormal{Im}}}

\DeclareMathOperator*{\argmax}{arg\,max}
\DeclareMathOperator*{\argmin}{arg\,min}

\begin{document}
  \title{High-temperature self-energy corrections to x-ray absorption spectra}

  \author{Tun S. Tan}
  \email[]{tunshengtan@ufl.edu}
  \affiliation{Quantum Theory Project, Department of Physics, University of Florida, Gainesville, Florida 32611, USA.}
  
  \author{J. J. Kas}
  \affiliation{Department of Physics, University of Washington, Seattle, Washington 98195-1560, USA.}
   
  \author{S. B. Trickey}
  \affiliation{Quantum Theory Project, Department of Physics, University of Florida, Gainesville, Florida 32611, USA.}
  
  \author{J. J. Rehr}
  \affiliation{Department of Physics, University of Washington, Seattle, Washington 98195-1560, USA.}
  \date{August 12, 2022}
  
  \begin{abstract}
    Effects of finite-temperature
    quasiparticle self-energy corrections to x-ray absorption spectra
    are investigated within the finite-temperature quasiparticle local density GW approximation up to  temperatures $T$ of order the Fermi temperature. To
    facilitate the calculations, we parametrize the quasiparticle
    self-energy  using low-order polynomial fits. We show that temperature-driven decrease in the electron lifetime substantially broadens the spectra in the near-edge region with increasing $T$. 
    However, the quasiparticle shift is most strongly modified near the onset
    of plasmon excitations.
  \end{abstract}


  \maketitle

    \section{Introduction}
        Calculations of x-ray absorption spectra (XAS) at finite temperature
        (FT) have been carried out routinely in recent years. These studies range from relatively low temperatures up to a few hundred K (LT), 
        to the
        warm dense matter (WDM) regime at high-temperatures (HT), where $T$
        is  of order the Fermi temperature $T_F$ \cite{PEYRUSSE2008, RECOULES2009, CHO2011, DORCHIES2015,
          ENGELHORN2015, OGITSU2018, BOLIS2019, JOURDAIN2020, ZI2021}. A
        standard FT approach is to apply Fermi's golden rule with initial and
        final states calculated using conventional density functional theory
        (DFT) and Fermi occupation factors.
        Since DFT is a ground state theory, 
        FT quasiparticle corrections to DFT
        are essential  for HT excited state calculations \cite{FALEEV2006, HOLLEBON2019}.
        However, even at extreme temperatures,
        e.g., many thousands of K, self-consistent field calculations of the
        core-hole state have sometimes been done with ground state
        exchange-correlation functionals $\varepsilon_{xc}[\rho]$ \cite{RECOULES2009, CHO2016, DENOEUD2014, JOURDAIN2020}. This
         ground-state approximation can be unreliable in that its validity
        depends strongly on the system state  
        and its properties.  Some  properties
        are only weakly sensitive to the temperature dependence of
        exchange and correlation, at least for low temperatures 
        well below the WDM
        regime. Others, such as the electrical conductivity and x-ray absorption spectra (XAS), are strongly temperature
        dependent \cite{Karasiev2016, KDT16}.  
        Nevertheless, the use of temperature-dependent 
        free-energy exchange-correlation functionals $f_{xc}[\rho,T]$ \cite{IIT1987, IIT2017, TANAKA2016,PDW2000, KSDT2014, GDTTFB2017,KDT16} alone ignores effects like inelastic losses. For example,
        the electron inelastic scattering effect, which results in an energy-dependent broadening, is often   included via a post-processing step by convolution
        of the absorption cross-section using an empirical model (e.g. Seah-Dench formalism \cite{XSPECTRA2013,SEAHDENCH1979}), or the imaginary part of the self-energy \cite{CUSHING2018}.
        However, the finite temperature dependence of the energy-dependent broadening is typically neglected.
    
        Another common approach for XAS calculations has been the use of
        the real-space multiple scattering (RSMS) method, which is also referred to as
        the real-space Green's function (RSGF) method \cite{REHR2000}. This approach
        is the real-space analog of the Korringa-Kohn-Rostoker (KKR) approach
        \cite{KORRINGA1947,KOHN1954,DUPREE1961,BEEBY1967,MORGAN1966}.  The
        method treats excited quasiparticle states via an energy-dependent self-energy, and
        also takes into account the dynamic response of the system to the suddenly
        created core-hole. The self-energy can be viewed as an
        energy-dependent, non-local analog of the exchange-correlation
        potential in DFT \cite{Casida1995}.  The FT generalization of the
        self-energy     can be done formally via the
        Matsubara formalism. For example, \citeauthor{BENEDICT2002} used the
        approach to investigate the effect of $T$ on the spectral function in
        jellium and aluminum, e.g., on optical properties of solid-density
        Al. Alternatively, as discussed by Kas et al. \cite{KAS2017}, the FT
        self-energy can be calculated using a generalization of the Migdal
        approximation \cite{ALLEN1983}, analogous to the FT $GW$ approximation
        of Hedin \cite{HEDIN1965}.
    
        Our main goal in this work is to discuss the effects of the
        FT GW self-energy on XAS, an approach that heretofore has not been
        explored in detail \cite{TAN2021}.
        In particular, to facilitate the calculations we introduce a
        parametrization of the quasiparticle FT GW self-energy within the
        $G_0 W_0$ scheme \cite{MartinReiningCeperley2016}.  As illustrations
        we apply the approach to the XAS for two systems with $T$ up to 10
        eV (i.e. $T$ of order $10^5$ K). Our calculations demonstrate that thermal broadening due to the imaginary part of the self-energy
        are significant above $T
        \approx 1$ eV. Although lattice vibrations also are strongly
        temperature dependent, that behavior is dependent on the lattice
        temperature $T_l$ which can differ from the electronic temperature $T$
        in non-equilibrium states, as discussed in a previous
        work \cite{TAN2021}.
    
        The remainder of the paper is organized as follows.  Section
        \ref{section:review}.\ provides an brief overview of the real-space Green's
        function approach to XAS and its dependence
        on the self-energy $\Sigma$.  In Section \ref{section:result},
        we highlight the FT corrections to XAS with a few examples and 
        in Section \ref{section:conclusion}, we present a brief summary
        and conclusions.  Throughout we use Hartree atomic units
        $q_e = \hbar = m = 1$, with $q_e=e$  the electron charge.  Thus
        energies are in Hartree and distances in Bohr,
        unless otherwise noted. For temperature we use either K or eV, with 1
        eV $\approx$ 11,604 K.  Electron densities are expressed in Wigner-Seitz
        radii $r_s = (3/4\pi \rho)^{1/3}$.  
    
    \section{Theory Summary}
        \label{section:review}
        \subsection{Finite-temperature X-ray Absorption}
        Formally the zero temperature X-ray absorption cross section
        is defined via Fermi's golden rule as
        \begin{eqnarray}
            \sigma(\omega) = 4\pi^2\frac{\omega}{c} \sum_{i,f} |\langle \Psi_i | \hat{\xi}\cdot \mathbf{R} | \Psi_f \rangle|^2 \delta_{\Gamma}(\omega + E_i -E_f),
        \end{eqnarray}
        where $|\Psi_i\rangle$ and $|\Psi_f\rangle$ are the many-body
        initial and final states, $\hat{\xi}$ is the polarization of the
        incident photon, and $\mathbf{R}$ is the many-body position
        operator. Then within the single-particle (quasiparticle) approximation  with dipole interactions and the 
        sudden approximation, the zero temperature XAS becomes
    
        \begin{eqnarray}
          \label{eqn:xas}
          \sigma_s(\omega) = 4\pi^2 \frac{\omega}{c}   \sum_{i,f} |\langle i |  {d} | f \rangle|^2 \delta_{\Gamma}(\omega + \varepsilon_i -\varepsilon_f),
        \end{eqnarray}
        where $\varepsilon_i$ and $\varepsilon_f$ are the energies of the
        quasiparticle initial $|i\rangle$ and final $|f\rangle$ levels and many-body shake-up factors $S_0^2 \approx 1$ are ignored.  The
        $\delta_\Gamma$ factor denotes a Lorentzian of width $\Gamma$ which
        includes both quasiparticle and core-hole lifetime broadening.
        Here, the transition operator $ {d=\hat\xi\cdot {\bf r}}=$ is the single-particle
        electric dipole operator.  
        The one-particle states $|i\rangle$ and $|f\rangle$ can be obtained from
        Hartree-Fock theory or Kohn-Sham DFT.  For the treatment via DFT,
        see  e.g., Refs.\ \citenum{OANA2013,TAILLEFUMIER2002,GOUGOUSSIS2009}.
    
        For x-ray
        absorption, the number of final states $|f\rangle$ required to compute
        the dipole matrix element has an impact on the computational efficiency of
        evaluating Eq.\ (\ref{eqn:xas}).  The present work uses the RSMS approach to alleviate  this  
        bottleneck. In RSMS, we replace the summation over the
        final states $|f\rangle$ with the retarded single-electron Green's
        function $G(\omega)$ in a basis of local site-angular momentum states
        $|Lj\rangle$ \cite{REHR2000},  
        \begin{eqnarray}
          G^{jj'}_{LL'}(\omega) = \sum_f \frac{\langle Lj|f\rangle\langle f| L'j'\rangle}{\omega-\varepsilon_f + i\eta} \; .
        \end{eqnarray}
        In this expression, $j$ is the index of a given site $\mathbf{R}_j$
        and $L=(l,m)$ are the angular momentum quantum numbers.  The initial
        states $|i\rangle$ are calculated with the ground state Hamiltonian $H =  p^2/2 + v(r)$ while
        the final states $|f\rangle$ are described by the quasiparticle Hamiltonian $H' =
        p^2/2 + v_f(r) + \Sigma(r,E)$, where $v(r)$ is the self-consistent one-electron Hartree potential,
        $v_f$ is the final state one electron Hartree potential in the presence of a
        screened core hole, and $\Sigma$ is the dynamically screened
        quasiparticle self-energy discussed in detail below.
            
        The imaginary part of the quasiparticle self-energy $\Sigma$ accounts for 
        the mean free path of the photoelectron.
        Within the quasiparticle local density approximation (QPLDA) \cite{SHAM1966}, the self-energy is given by \cite{MUSTRE1991}
        \begin{eqnarray}
          \Sigma(\textbf{r}, E,T=0) &=& v^{LDA}_{xc}(\rho(\textbf{r})) + \Sigma_{GW}(\rho(\textbf{r}), E, T=0)\nonumber\\
                                & & - \Sigma_{GW}(\rho(\textbf{r}), E_F, T=0) \; .
        \end{eqnarray}
        Here $\Sigma_{GW}$ is the $GW$ self-energy calculated at the $G_0W_0$ level of refinement, that is, without self-consistent iteration of $G$ or $W$ \cite{HEDIN1965}.
        For simplicity from here onward, we drop the spatial dependence $\textbf{r}$. 
        
        For the FT generalization, we replace the $T=0$ $GW$ self-energy with the finite-temperature $GW$ self-energy, $\Sigma_{GW}(T)$, and
        introduce $T$-dependent Fermi occupation numbers, $f(\varepsilon) = 1/[\exp{\{\beta(\varepsilon-\mu)\}} + 1]$ for the initial and
        final states in Eq.\ (\ref{eqn:xas}).
        In addition, the ground state exchange-correlation
        potential $v^{LDA}_{xc}$ is replaced by its FT generalization
        $v^{LDA}_{xc}(T)$. Thus, the finite-temperature QPLDA self-energy is
        \begin{eqnarray}
          \Sigma(E, T) &=& v^{LDA}_{xc}(\rho,T) + \Sigma_{GW}(\rho,E, T) \nonumber\\
                                & & - \textnormal{Re}\ [\Sigma_{GW}(\rho,\mu_T, T)]
        \end{eqnarray}
        Lastly, by using $G^{jj'}_{LL'}(\omega)$ in Eq.\ (\ref{eqn:xas}) in place of the sum over final states $|f\rangle$, the FT quasiparticle cross section can be re-expressed as:\cite{ANKUDINOV1998}
        \begin{eqnarray}
          \sigma_{qp}(\omega) &=& - 4\pi^2 \frac{\omega}{c} \im  \sum_{iLL'} \langle i | \hat{d} G^{00}_{LL'}(\omega+\varepsilon_i) \hat{d}^\dagger | i \rangle \nonumber\\
                              & & \times f(\varepsilon_i) \big[ 1 - f(\omega + \varepsilon_i) \big],
        \end{eqnarray}
        Here, we denote the absorbing atom by the index 0.
        
        \subsection{Finite-temperature Self-energy $\Sigma$}
            The finite-$T$ quasiparticle electron self-energy within the $GW$ approximation is defined formally \cite{ALLEN1983,MAHAN2000}  by the expression
            \begin{eqnarray}
              \Sigma^M_{GW}(\bk,i\omega_m) &=& -\frac{1}{\beta} \int \frac{d^3\bq}{(2\pi)^3} \sum_{n=-\infty}^{\infty} G_0^M (\bk-\bq,i\omega_m-i\nu_n) \nonumber\\
                               & &      \times W^M(\bq,i\nu_n) \;.
            \end{eqnarray}
            Here $G_0^M$ is the one-electron Matsubara Green's function, $W^M=
            \epsilon^{-1} v$ is the screened Coulomb interaction, and
            $\omega_m=2(m+1)\pi k_B T$, $\nu_n=2n\pi k_B T$ are the Matsubara
            frequencies, where $\epsilon$ is the dielectric function and $v$ is
            the bare Coulomb potential. The screened interaction $W^M$ can be
            expressed in terms of its spectral representation as
            \begin{eqnarray}
              W^M(\bq, i\nu_n) &=& v(\bq) + \int_{-\infty}^{\infty} d\omega' \frac{D(\bq, \omega')}{i\nu_n - \omega' + i\eta\, \sgn(\omega')}
            \end{eqnarray}
            where $v(\bq)=4\pi/q^2$ is the bare Coulomb potential in Fourier
            representation, and $D(\bq,\omega) = (1/\pi) |\im\, W^M_c(\bq, \omega)|\,\sgn(\omega)$ is the anti-symmetric
            (in frequency) bosonic excitation spectrum. $W^M_c = W^M - v$ is the correlation part of the screened interaction. 
    
            Our choice of the electron gas dielectric function reflects a balance between the
            level of physics included and computational feasibility. Thus for
            simplicity, we use the random phase approximation (RPA), which is
            analogous to the FT generalization of the Lindhard function \cite{ARISTA1984},
            \begin{eqnarray}
              \epsilon(\bq, \omega, T) &=& 1 + 2v(\bq) \int \frac{d^3\bk}{(2\pi)^3} \frac{f(\varepsilon_{\bk-\bq})-f(\varepsilon)}{\omega - \varepsilon_{\bk-\bq} + \varepsilon_\bk + i\eta},
            \end{eqnarray}
            where $f(\varepsilon) = 1/[\exp{\{\beta(\varepsilon-\mu)\}} + 1]$ is
            the Fermi-Dirac occupation factor, and $\mu = \mu(T)$ is the chemical potential.
            The real part of $\epsilon(\bq, \omega, T)$ is obtained from the
            imaginary part via a Kramers-Kronig transform.
            From an analytic continuation to the real-$\omega$ axis,
            the FT GW retarded self-energy $\Sigma_{GW}$ is given by the Migdal
            approximation \cite{ALLEN1983}
            \begin{eqnarray}
              \label{eqn:ft_sigma}
              &&\Sigma_{GW}(\bk, \omega, T) = \Sigma_{X}(\bk, \omega, T) + \int_0^{\infty} d\omega' \int \frac{d^3\bq}{(2\pi)^3} D(\bq,\omega') \nonumber\\
                   &\times&  \bigg[ \frac{f(\varepsilon_{\bk-\bq}) + N(\omega')}{\omega+\omega'-\varepsilon_{\bk-\bq} + i\eta}
                      + \frac{ 1 - f(\varepsilon_{\bk-\bq}) + N(\omega')}{\omega-\omega'-\varepsilon_{\bk-\bq} + i\eta} \bigg],
            \end{eqnarray}
            where $\Sigma_{X}(\bk,\omega,T) = \int [d^3\bq/(2\pi)^3] f(\varepsilon_{\bk-\bq})v_\bq$ is the exchange part of self-energy and $N(\omega) = 1/[\exp\{\beta\omega\}- 1]$ is the Bose factor. The poles of the Green's function $G^M$ contribute to the Fermi occupations whereas the poles of the screened interaction $W^M$ contribute to the Bose factor.
    
\begin{figure}
                \includegraphics[width=0.45\textwidth]{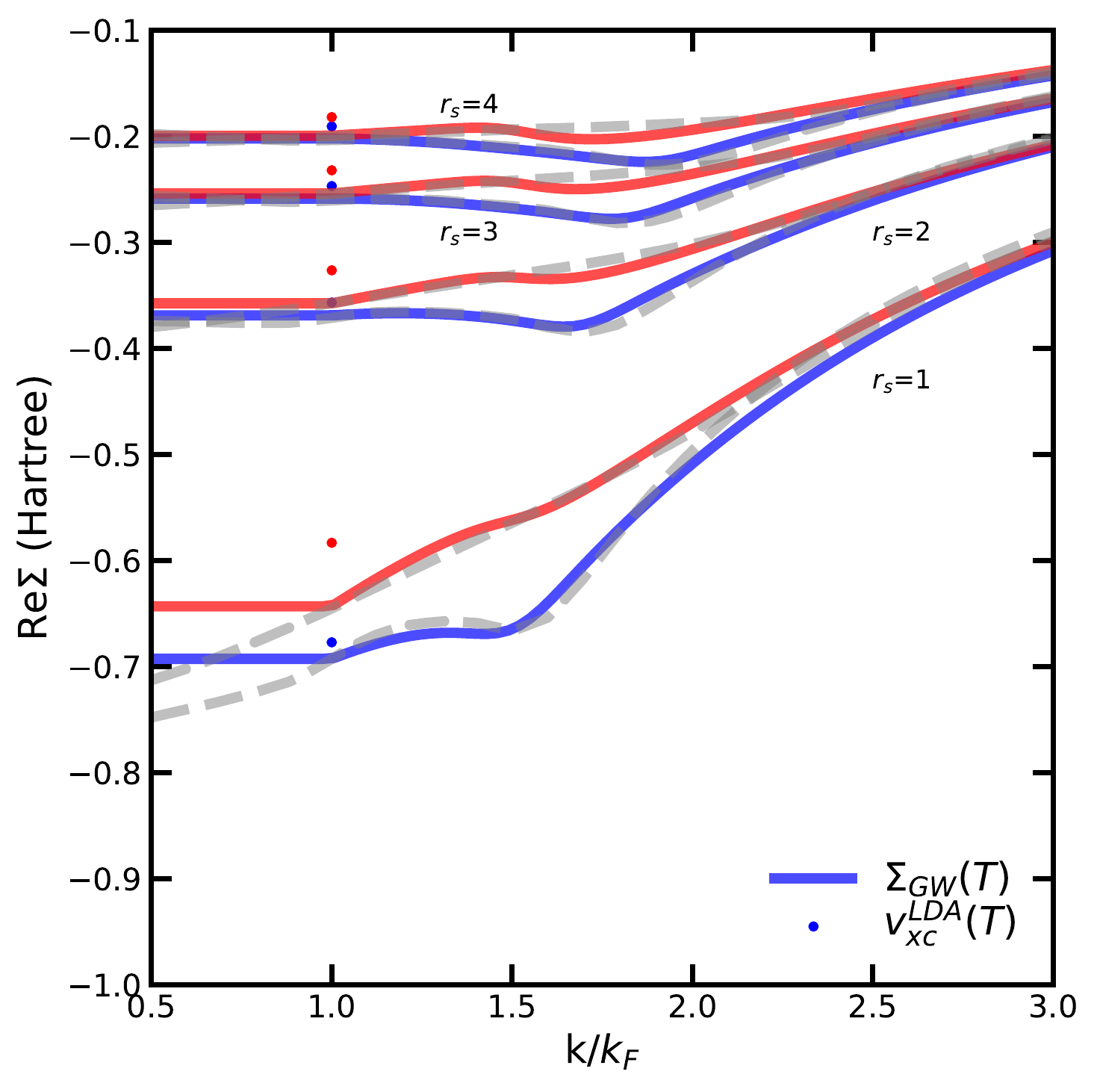}
                \caption{The real part of the $\Sigma_{GW}$ (dashed) and our parametrization (solid) for various densities: $r_s$ = 1.0, 2.0, 3.0, 4.0 at $T/T_F$ = 0.01 (blue) and 1.0 (red). For reference the LDA-$v_{xc}$(T) \cite{KSDT2014} values are denoted as dots at $k_F$.}
                \label{fig:gw_vs_ksdt}
            \end{figure}

            Calculations of the imaginary part of 
            $\Sigma_{GW}(\bk, \omega, T)$ involves a singe integral over the
            magnitude of $\bq$ but to obtained the real part
            we need to perform a Kramers-Kronig transform resulting in a double integral.
            In  typical RSGF XAS
            calculations \cite{FEFF10}, tens of thousands of self-energy evaluations
            are required.
            Thus that calculation of quasiparticle self-energy $\Sigma_{GW}( k,k^2/2,T)$ immediately
            becomes a computational
            bottleneck.
            To circumvent that difficulty, we model $\Sigma_{GW}(k,k^2/2,T)$ via low-order polynomial fits
            to numerical calculation of $\Sigma_{GW}$ based on Eq.\ (\ref{eqn:ft_sigma}), on a grid up to $T = 2 T_F$, where $T_F=E_F/k_B$.   The
            form of the fitting functions is described in detail in the Appendices. A
            comparison between the QPLDA $\Sigma_{GW}$ and the fits  is shown in
            Fig.\ \ref{fig:gw_vs_ksdt} for the homogeneous electron gas.
            For simplicity, we approximate the self-energy for levels $k < k_F$ with $v_{xc}(T)$ independent of $k$.    
            
    \section{Results and Discussion}
        \label{section:result}
        In this section, we present our results showing effects of the
        use of the FT $\Sigma_{GW}(T>0)$ instead of the zero $T$
          values $\Sigma_{GW}(T=0)$.
        Note that $\Sigma_{GW}(T=0)$  depends implicitly on the
        electronic temperature  $T$ through the self-consistent electron
        density at that temperature. The FT SCF calculations are carried out using an extension of the original RSMS code, which is now implemented  in  FEFF10 \cite{FEFF10, FEFF9}. For the FT exchange
        correlation-potential, we use the KSDT tabulation $v^{LDA}_{xc}(T)$ \cite{KSDT2014,KTD2019PRB} for both calculations.
        In the FEFF10 calculations, we compute the atomic part with the ground state exchange potential,  whereas we use $\Sigma_{GW}(T)$ for the fine structure.
    
        As noted in Ref. \cite{RECOULES2009, JOURDAIN2020}, corrections to the core energy levels are needed for $T \gtrsim 1$ eV. We compute the core level shifts using the
        all-electron full-potential linearized plane wave code FLEUR \cite{FLEUR, FLPAW1982, FLEUR2018}.
        Within these calculations, we ignore the explicit
        temperature dependence in the LDA exchange correlation functional \cite{PZ1981}.
        As a side note, FLEUR uses non-overlapping muffin-tin potentials whereas FEFF uses overlapping muffin-tin potentials.
    
\begin{figure}
          \includegraphics[width=0.45\textwidth]{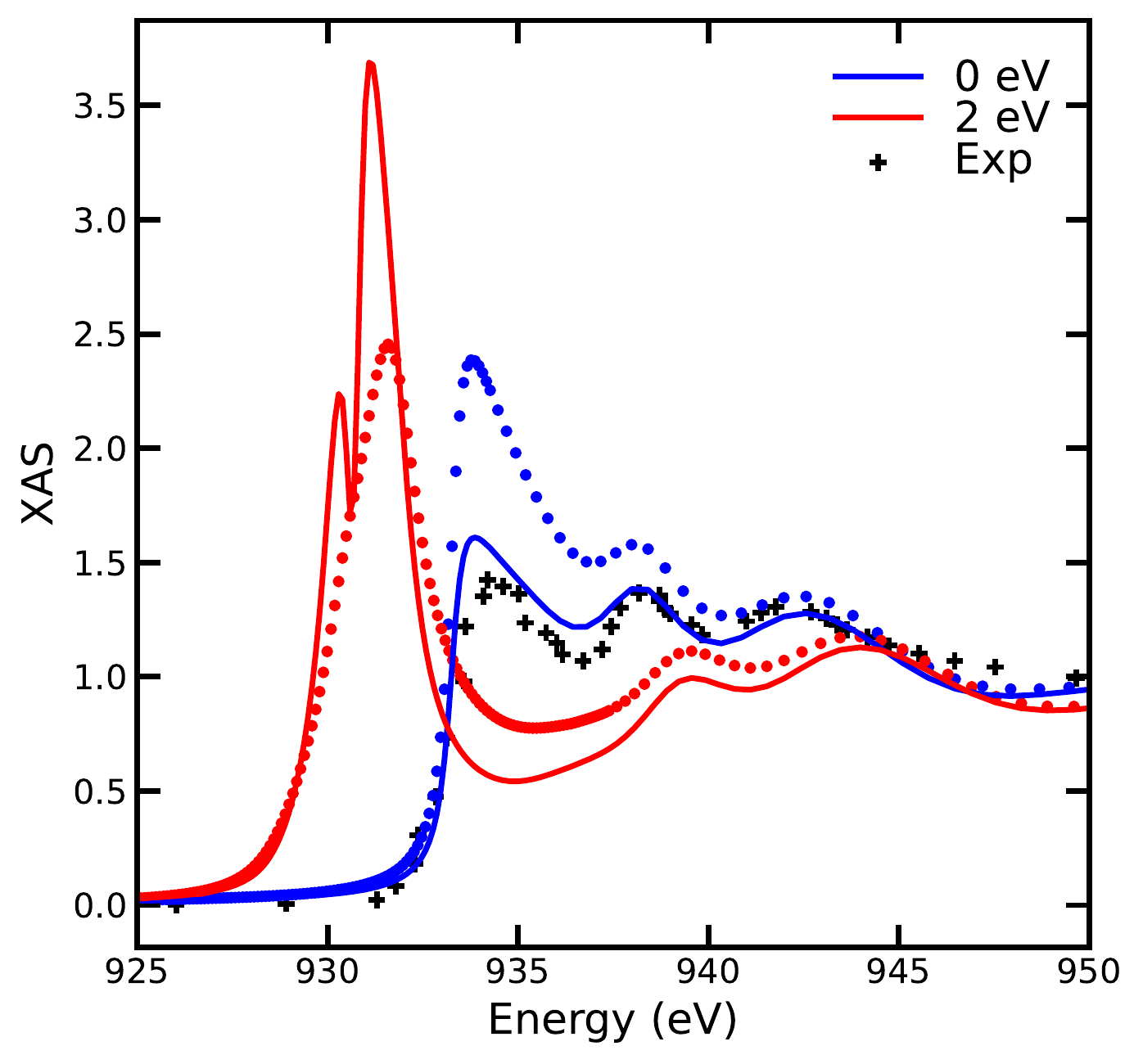}
          \caption{$\textnormal{L}_3$-edge XAS for Cu (lattice constant $a = 3.61$ \AA \cite{KITTEL2005}) at  electronic temperature $T=0$ eV and 2 eV.
          The solid curves denote the absorption with ground state atomic background while
          the dots represent the absorption with $\Sigma_{GW}(T>0)$ self-energy atomic background. The experimental measurement at ambient condition is shown as black cross.\cite{KIYONO1978}}
          \label{fig:atomic_background}
        \end{figure}

        In the RSMS formalism, the decomposition of the Green's function $G$
        into a central atom $G^c$ contribution and a multiple-scattering
        $G^{sc}$ contribution allows us to describe the XAS 
        in terms of the atomic background $\sigma_0$
        and the oscillatory fine structure $\chi$, i.e., $\sigma = \sigma_0 (1 + \chi)$.  For many XANES calculations, it is found
        that the atomic background matches the experimental results better
        when calculated without the self-energy corrections due to the
        overestimation of the exchange within the muffin-tin (MT) potential
        approximation \cite{REHR1994}.
        Fig.\ \ref{fig:atomic_background} shows the effect of using the ground-state potential versus the use of $\Sigma_{GW}(T)$ for the atomic background. 
        The pre-edge is dominated by the atomic background and thus is sensitive to the choice of exchange potential.
        The pre-edge amplitude is reduced by
        $\approx 28\%$ due to the temperature correction of $\Sigma_{GW}(T)$.
        More pump-probe experimental XAS measurements for
        $T \approx 10$ eV
        are required to validate the FT self-energy effect for the atomic background.
        Nonetheless, the FT self-energy corrections are important for the description of HT fine-structure.
    
\begin{figure}
            \includegraphics[width=0.45\textwidth]{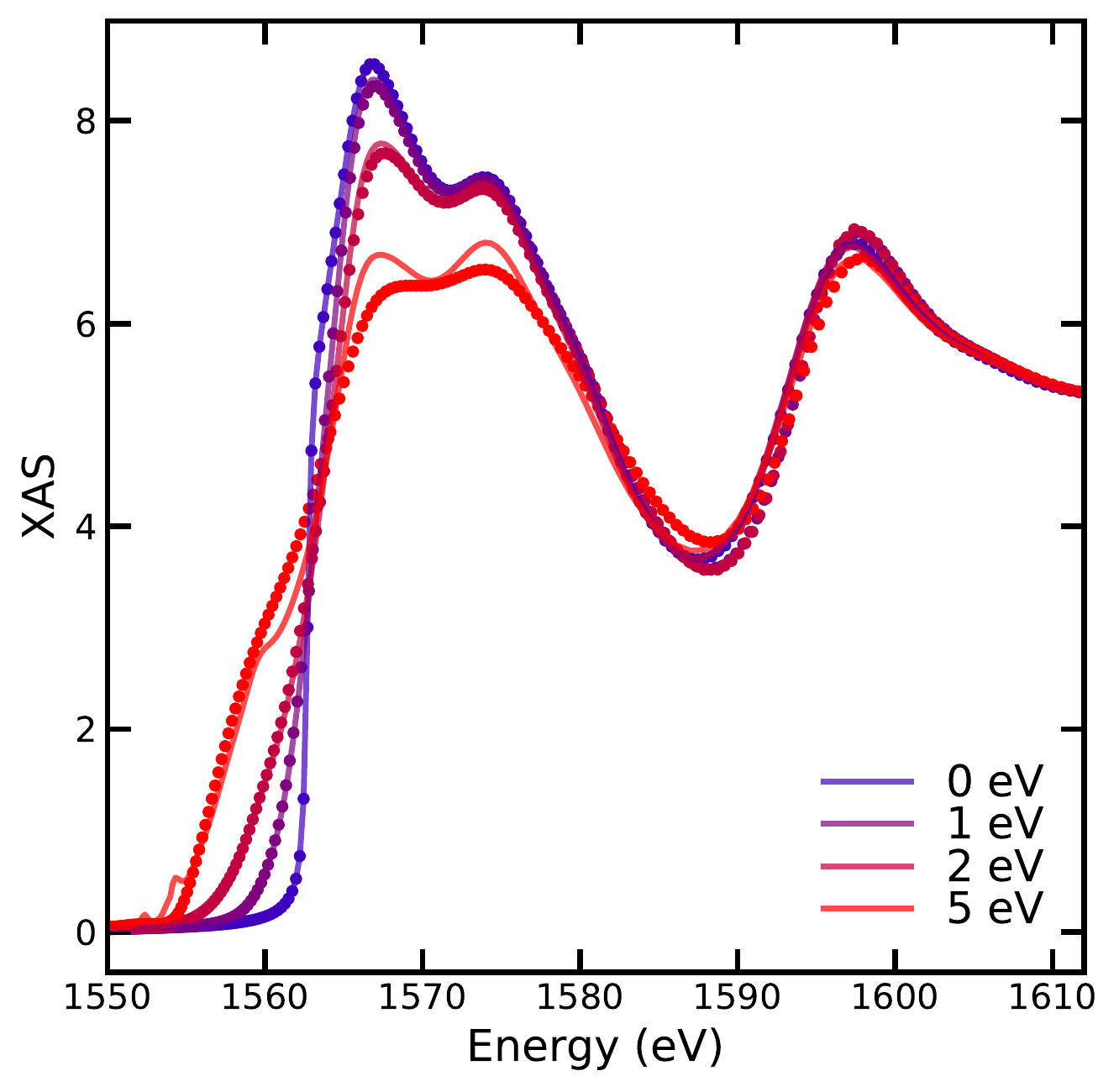}
            \caption{
            $\textnormal{K}$-edge XAS for fcc aluminum ($a = 4.05$ \AA) using different self-energies:
            $T$-independent GW self-energy $\Sigma_{GW}(T=0)$ (solid curves) and $T$-dependent GW self-energy $\Sigma_{GW}(T>0)$ (dots).}
            \label{fig:1}
          \end{figure}

        As a first example, we consider the FT K-edge x-ray absorption
        near-edge spectrum (XANES) for aluminum (fcc Al, lattice constant $a = 4.05$ \AA \cite{KITTEL2005}).  Al is a
        prototypical nearly-free electron system in the sense that
        the electronic density
        of states (DOS) in the conduction band has a nearly square root like
        dispersion at the bottom of the band. 
        Fig.\ \ref{fig:1} shows the comparison of the Al K-edge
        spectrum at different temperatures including or excluding explicit electronic T-dependent effects in the self-energy, namely, using $\Sigma_{GW}(T=0)$ or $\Sigma_{GW}(T>0)$.
        When restricting the temperature T solely to that which is introduced through the density ($\Sigma_{GW}(T=0)$  case), we observe the broadening of the edge and no shift in the edge position. The shift in the core levels compensates for the shift in the valence density. On the other hand, for $T$ up to about 1 eV, the finite
        temperature self-energy correction is negligible, and the temperature-independent electron self-energy model is a good
        approximation. However, as temperature grows to  order $\approx$ 10 eV,
        the fine structures are smoothed by the large broadening ($\approx$ 3
        eV) associated with shortened electronic excitation lifetime.  The
        shift in quasiparticle shift due to the finite temperature self-energy
        correction is small for the near-edge region, and only becomes
        significant between 10 eV and 20 eV above the chemical potential.
        
\begin{figure}
            \includegraphics[width=0.45\textwidth]{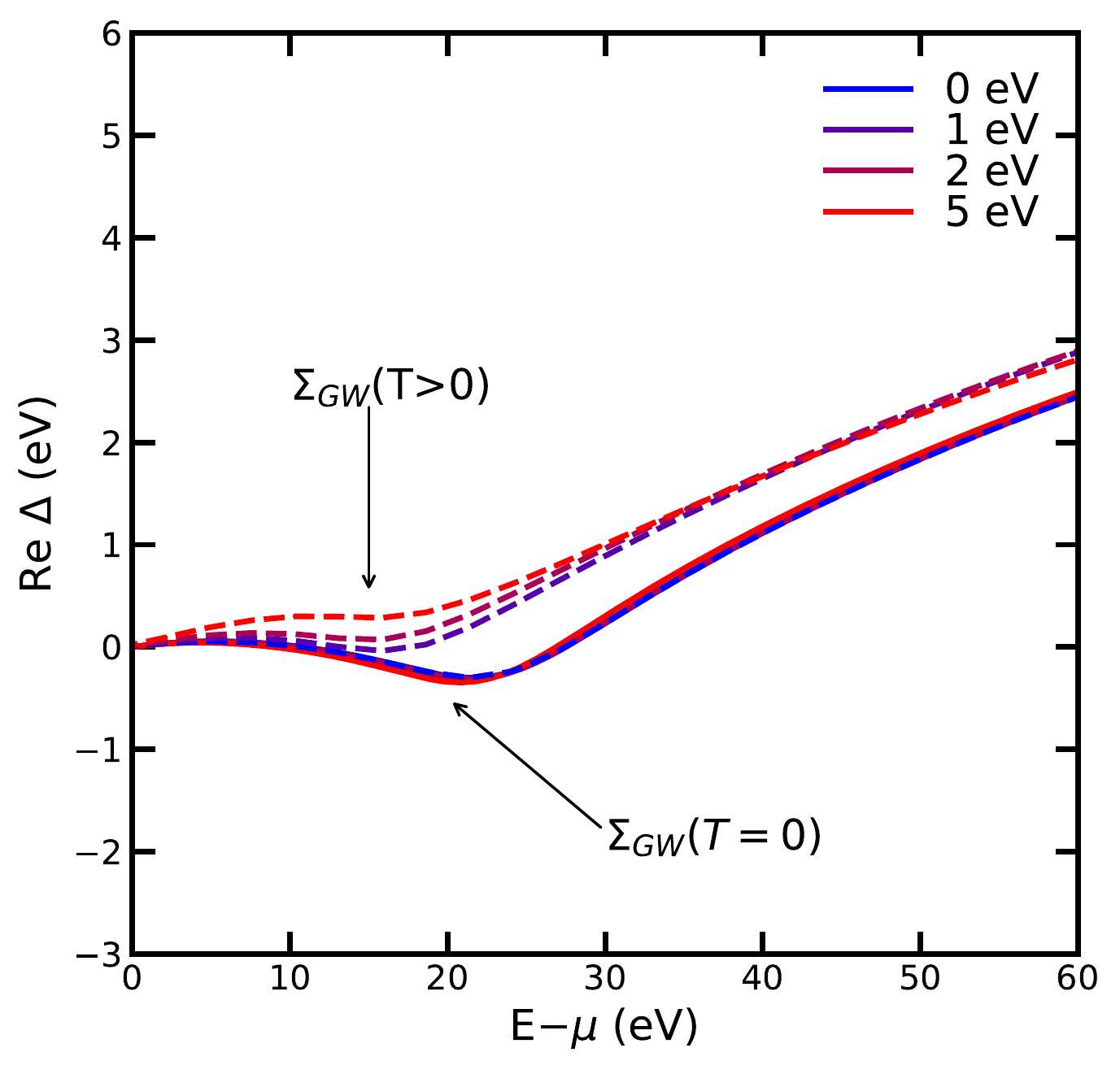}
            \caption{The quasiparticle corrections Re $\Delta_k$ for aluminum at temperatures $T=0, 1, 2$ and $5$ eV.
            The calculations used KSDT $v_{xc}(T)$ and different self-energies: $\Sigma_{GW}(T=0)$ (solid), and $\Sigma_{GW}(T>0)$ (dashed).}
            \label{fig:1a}
          \end{figure}

\begin{figure}
            \includegraphics[width=0.45\textwidth]{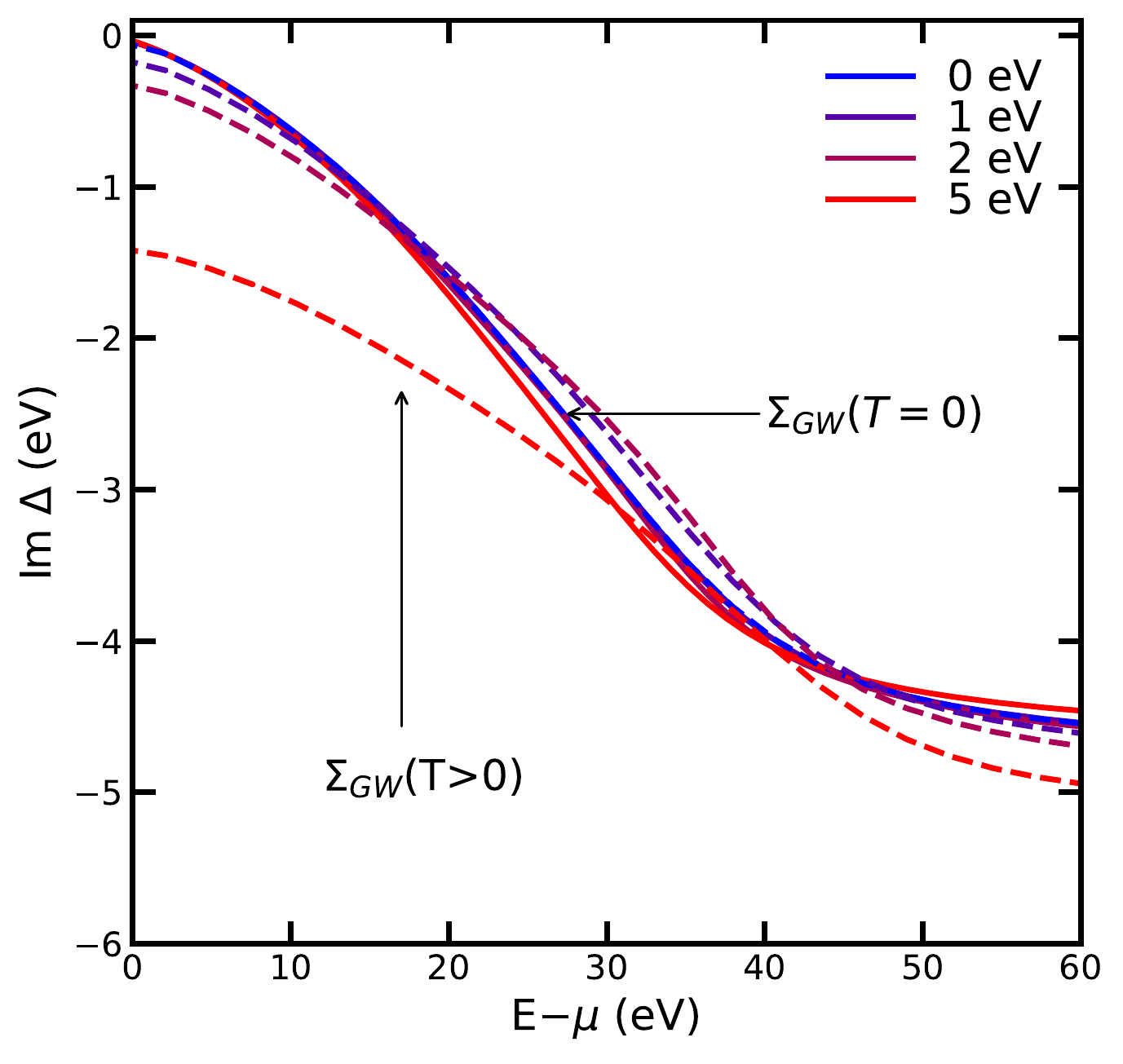}
            \caption{The quasiparticle corrections Im $\Delta_k$ for aluminum at temperatures $T=0, 1, 2$ and $5$ eV.
            The calculations used KSDT $v_{xc}(T)$ and different self-energies:  $\Sigma_{GW}(T=0)$ (solid), and $\Sigma_{GW}(T>0)$ (dashed).}
            \label{fig:1b}
          \end{figure}

        As a further illustration of the explicit $T$-dependence of the FT
        quasiparticle self-energy, we compute the quasiparticle energy
        correction $\Delta_k = \varepsilon_{qp} - \varepsilon_{k}$. Here, the quasiparticle energy $\varepsilon_{qp}$ is the solution to $\varepsilon_{qp}(k') = \varepsilon_{k} + \Sigma_{GW}(k', \varepsilon_{qp}(k'))$ and $k = \sqrt{2(E-\mu)}$ is the photoelectron wavenumber. We compare
        the real part of $\Delta$ in Fig.\ \ref{fig:1a} and imaginary part in
        Fig.\ \ref{fig:1b} for different self-energies at the interstitial density:
        $\Sigma_{GW}(T=0)$ and $\Sigma_{GW}(T>0)$. 
        Note that the  real part of $\Delta$
        shows a strong temperature dependence between 5 eV and 20 eV near the
        plasmon onset.  For the imaginary part of $\Delta$, the broadening effect becomes important above $T$ = 1 eV.
    
\begin{figure}
          \includegraphics[width=0.45\textwidth]{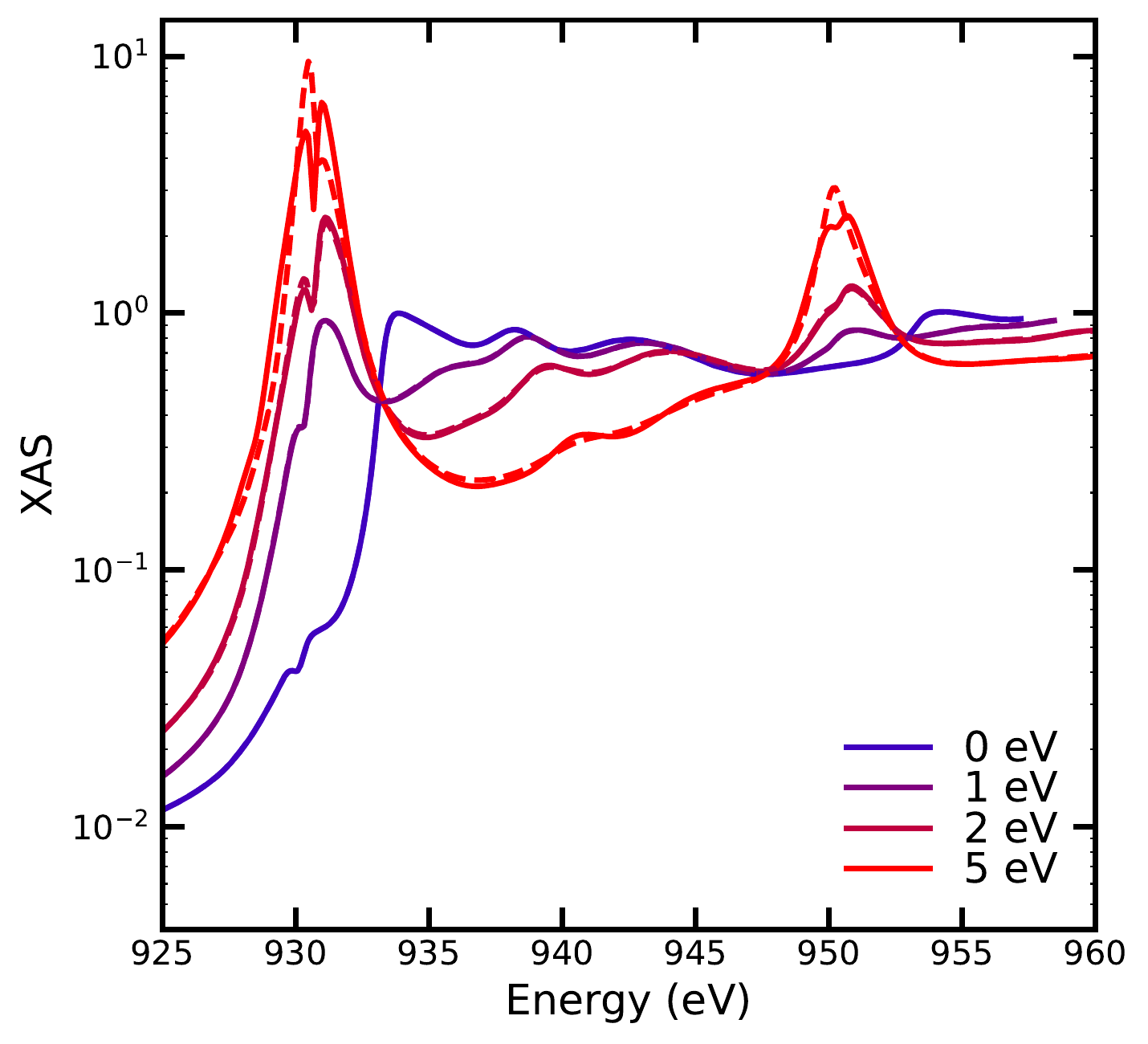}
          \caption{$\textnormal{L}_{3,2}$-edge XAS for Cu ($a = 3.61$ \AA) finite electronic temperature $T=0, 1, 2$ and $5$ eV, where the structure reflects that of the unfilled $d$-bands.
          The solid curves denote the $\Sigma_{GW}(T=0)$ self-energy results while
          the dashes represent the $\Sigma_{GW}(T>0)$ self-energy results.}
          \label{fig:copper_l_edge}
        \end{figure}

        As a second example, we present results for a noble transition metal
        (fcc copper, lattice constant $a = 3.61$ \AA \cite{KITTEL2005}) for which the $d$-bands are essentially full. Unlike the
        K-edge, the L-edge probes the highly-localized $d$-bands of Cu near
        the chemical potential. At high temperatures, the pre-edge peak
        increases in amplitude due to the decreasing $d$-state
        occupation \cite{JOURDAIN2020, TAN2021}. Fig.\ \ref{fig:copper_l_edge}
        shows the $\textnormal{L}_{3,2}$-edge XAS up to $T=5$ eV.  The
        temperature dependence of $\textnormal{Im}\ \Sigma_{GW}(T)$ results in
        changes to pre-edge peaks at
        $T \gtrapprox 2$ eV. Consequently, 
        the estimation of temperature based on the pre-edge area method or
        direct spectrum fitting will deviate more from the $\Sigma_{GW}(T=0)$
        model as temperature increases.
    
    \section{Summary, Conclusions, and Outlook}
        \label{section:conclusion}
    
        Our parametrization of the FT GW electron self-energy enables
        efficient calculations of XAS at finite $T$, from LT of a few hundred K,  up to the WDM regime
        with $T$ at least 10 eV. Our strategy uses the QPLDA $G_0W_0$ level of
        refinement for the self-energy, with the RPA dielectric function in conjunction with the
        KSDT finite-$T$ LDA exchange-correlation functional.  Specifically,
        the FT self-energy for a system is approximated using the uniform
        electron gas with density equal to that of the local density. This is a significant
        simplification, as direct calculations using the exact loss function
        of the system for the entire energy range of typical XAS experiments would be
        computationally formidable.
        A finite-temperature 
        SCF procedure for the XAS calculations is carried out in the complex energy plane in terms of
        the FT one-electron Green’s function. The procedure includes the FT
        exchange-correlation potential, approximated here by  the KSDT
        parametrization.  Important FT XAS effects include
          the smearing of the absorption edge and the presence of
        peaks in the spectrum below the $T = 0$ K Fermi energy. The FT
        exchange-correlation potential has only a small effect on XAS at low temperatures $T \ll
        T_{F}$ compared to the effect of Fermi smearing. The FT self-energy is also
        important for XAS, accounting for both temperature dependent 
        shifts and final-state
        broadening.
        To illustrate its efficacy, 
        the approach was applied to calculations of XANES for
        crystalline Al and Cu at normal density.  Above $T>1$ eV, the
        fine structures experience substantial broadening in the K-edges, corresponding to a 
        reduction of the quasiparticle lifetime with  increasing $T$.
        
        Going forward, a computationally efficient approximation beyond 
        the uniform electron gas dielectric function would be to use a 
        many-pole model \cite{KAS2007, KAS2009}, which is   an extension of
        the Hedin-Lundqvist single plasmon-pole model \cite{LUNDQVIST1967,HEDIN1971}. {\it Ab-initio} dielectric
        functions  also can be obtained from modern electronic structure
        codes. Such a finite-temperature generalization of the many-pole model is currently
        under development.

    \acknowledgments
        The contributions from  JJK and JJR are supported
        by the Theory Institute for Materials and Energy Spectroscopies (TIMES) at SLAC funded by the U.S. DOE, Office of Basic Energy Sciences, Division of Materials Sciences and Engineering under contract EAC02-76SF0051.
        TST and SBT are supported by DOE grant DE-SC0002139.
    \appendix
    \section{Model for $\textnormal{Re}\ \Sigma_{GW}$}
\begin{figure}
          \includegraphics[width=0.45\textwidth]{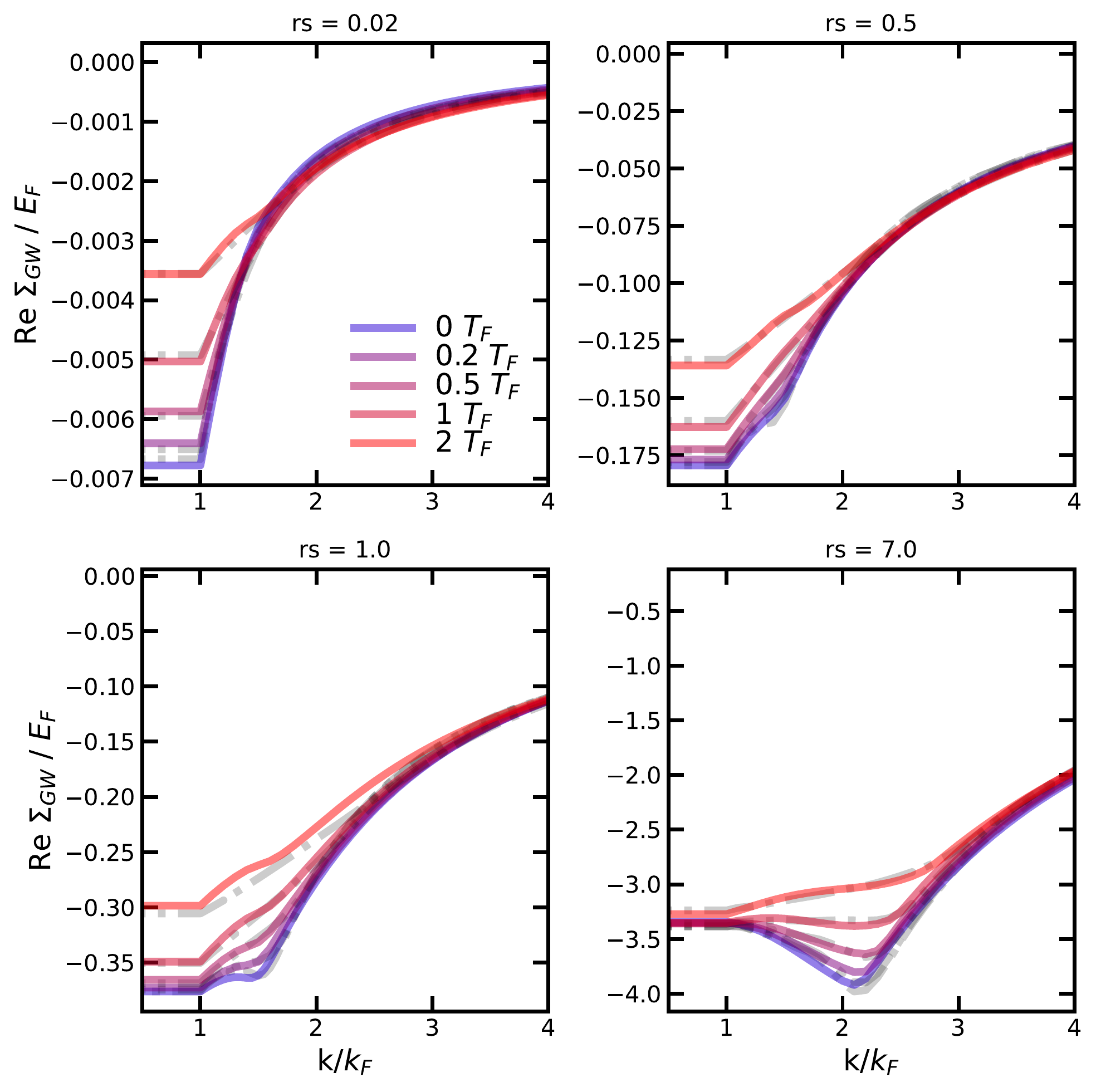}
          \caption{\label{fig:real_fit} Our parametrization (solid) to the real part of $\Sigma_{GW}(T)$ (gray, dash-dotted) for $r_s$ = 0.02, 0.5, 1.0 and 7.0 (blue to red).}
        \end{figure}

        The real part of the finite temperature GW self-energy, $\textnormal{Re}\ \Sigma_{GW}(T)$, is parametrized using low-order polynomials. 
        The parametric variables are the Wigner-Seitz radius $r_s$, reduced momentum $x = k/k_F$ and reduced temperature $t = T/T_F$. Below the variable $X$ represents the array denoted $X=(r_s,x,t)$.
        \begin{eqnarray}
          \frac{ \textnormal{Re}\ \Sigma_{GW}(X)}{E_{F}} &=& \begin{cases} 
            \alpha_0(t) + \alpha_1(t)\beta_1(r_s) x \\
            + \alpha_2(t)\beta_2(r_s) x^{3/2} \\
            + \alpha_3(t)\beta_3(r_s) x^2 \\
            + \alpha_4(t)\beta_4(r_s) x^{5/2},x < \kappa(r_s, t) \\
            \\
            \tilde{\alpha}_1(t) \tilde{\beta}_1(t)/x^{-1} \\ +\tilde{\alpha}_2(t) \tilde{\beta}_2(t)/x^{-2} \\
            +\tilde{\alpha}_3(t) \tilde{\beta}_3(t)/x^{-3},  x \geq \kappa(r_s, t)
          \end{cases}
        \end{eqnarray}
        Here $\alpha_i(t) = \alpha_{i1} + \alpha_{i2} t + \alpha_{i3} t^2$ and $\tilde{\alpha}_i(t)$ has the same form. Similarly
        $\beta_i(r_s) = \beta_{i1} r_s + \beta_{i1} r_s^{3/2} + \beta_{i1} r_s^{2}$ and $\tilde{\beta}_i(r_s)$ has the same form.
        The function $\kappa$ is defined as:
        \begin{eqnarray}
            \kappa(r_s, t) &=& \bigg(1 + \tanh \bigg[ a_1 r_s - \alpha_1(t) \bigg] \bigg) \nonumber\\
                        & &\times \bigg(a_2 \log\big[ r_s \big]^2 + \alpha_2(t) \bigg) + \alpha_3(t)
        \end{eqnarray}
        where it is fitted to the position of the cusp defined as
        \begin{eqnarray}
          \textnormal{cusp} &=& \begin{cases} 
                        
                        \min(1.5, \argmin_{x}\textnormal{Re}\ \Sigma_{GW}),\ 0.2 < r_s \leq 5\\
                        \min(1.5, \argmax_{x}\partial_x \textnormal{Re}\ \Sigma_{GW}),\ r_s > 5\\
                        1,\ r_s \leq 0.2\\
            \end{cases}
        \end{eqnarray}
        The resulting $\textnormal{Re}\ \Sigma_{GW}$ parametrization is shown in Fig.\ \ref{fig:real_fit}.
        The absolute mean error for our parametrization is $\approx$ 0.007 $E_F$.
    
    \section{Model for $\textnormal{Im}\ \Sigma_{GW}$}
        The imaginary part of the finite temperature GW self-energy, $\textnormal{Im}\ \Sigma_{GW}(T)$, is parametrized by
        \begin{eqnarray}
            \frac{\textnormal{Im}\ \Sigma_{GW}(X)}{E_F} &=&
            \begin{cases}
                \eta_1(r_s) k t +  \eta_2(r_s) k t^{\frac{3}{2}} \\
                + \eta_3(r_s) k^2 t +  \eta_4(r_s) k^2 t^{\frac{3}{2}}\\
                + \eta_5(r_s) k^{\frac{3}{2}} t +  \eta_6(r_s) k^{\frac{3}{2}} t^{\frac{3}{2}} \\
                + \eta_7(r_s) k +  \eta_8(r_s) k^2  \\
                + \eta_9(r_s) k^{\frac{3}{2}} \\
                + \frac{\textnormal{Im}\ \Sigma_{GW}(r_s, x=1, t)}{E_F}, x < \lambda(r_s, t) \\
                \\
                \sigma\sum_{i=1}^{6} \tilde{\eta}_i(r_s) \beta_i(t) k^{-i}, x \geq \lambda(r_s, t)
            \end{cases} 
          \label{eqn:im_model}
        \end{eqnarray}
        where $\eta_i(r_s), \tilde{\eta}(r_s) = \eta_{i1} r_s + \eta_{i2} r_s^{3/2} + \eta_{i3} r_s^2$ and $\sigma$ is the standard deviation of the data points used.
        The function $\lambda(r_s, t)$ is given by
        \begin{eqnarray}
          \lambda(r_s, t) &=& \bigg(1 + \tanh \bigg[ \frac{r_s - \frac{1}{2}}{p_1} \bigg] \bigg) \nonumber\\
                         & & \times \bigg(p_2 r_s + \eta_1(t) \bigg) + \eta_2(t)
          \label{eqn:im_switch}
        \end{eqnarray}
        and Im $\Sigma$ at the Fermi level, $k_F$, is parametrized by:
        \begin{eqnarray}
          \frac{\textnormal{Im}\ \Sigma_{GW}(r_s, x=1, t)}{E_F} &=& \nu_1(t) r_s^{\frac{1}{2}} +  \nu_2(t) r_s\nonumber\\
          & & + \nu_3(t) r_s^{\frac{3}{2}} +  \beta_4(t) r_s^{\frac{5}{2}}
          \label{eqn:im_kf_model}
        \end{eqnarray}
        where $\nu_i(t) = \nu_{i1} t + \nu_{i2} t^{3/2} + \nu_{i3} t^2$. The fittings of Eq.\ (\ref{eqn:im_model}), Eq.\ (\ref{eqn:im_switch}) and Eq.\ (\ref{eqn:im_kf_model})  are done for two temperature regions: $ t < 0.5$ and $t \geq 0.5$,
        and two density regions: $r_s < 0.2$ and $r_s \geq 0.2$. 
    
\begin{figure}
          \includegraphics[width=0.45\textwidth]{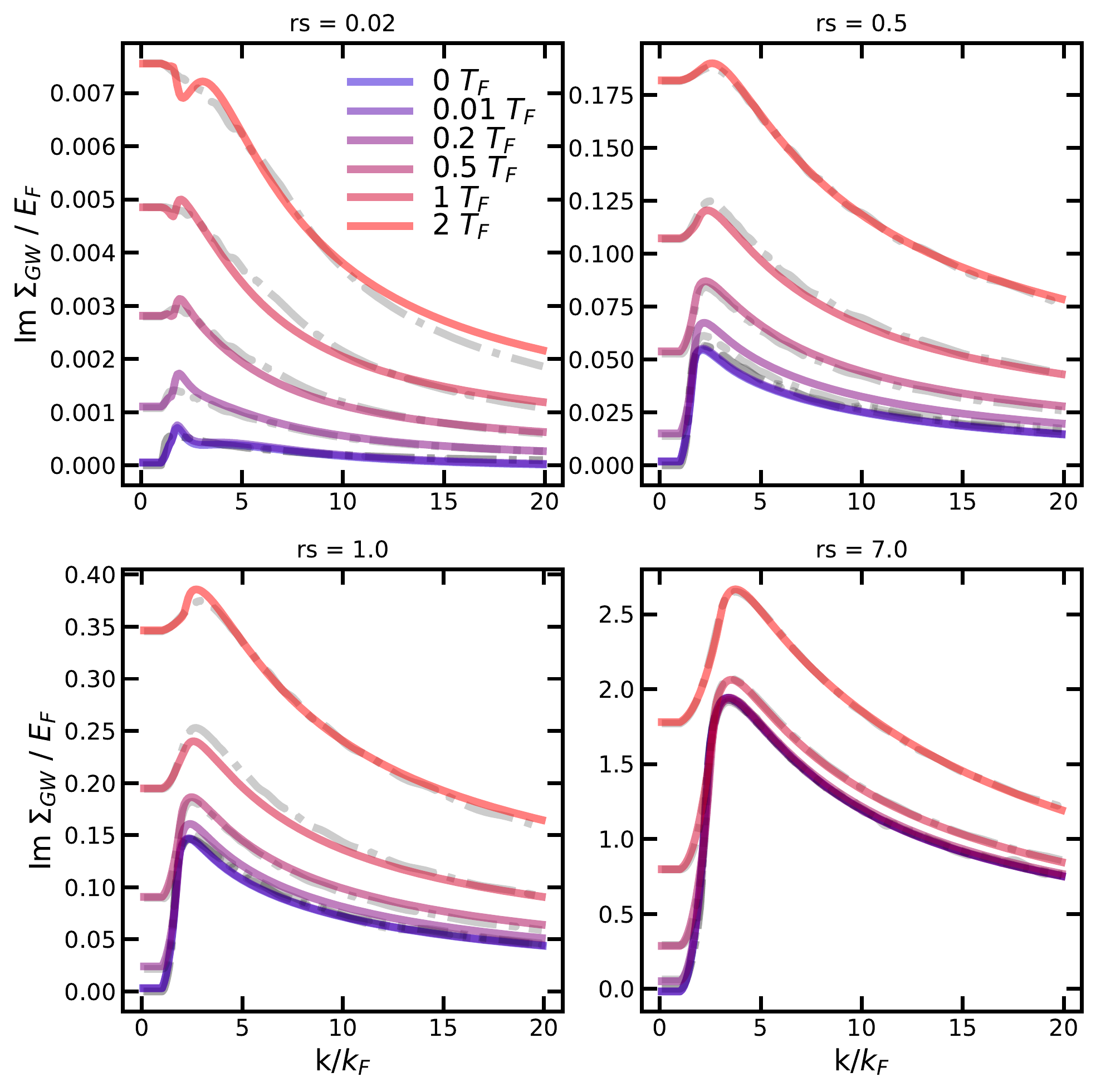}
          \caption{\label{fig:imag_fit}  Our parametrization (solid) to the imaginary part of $\Sigma_{GW}(T)$ (gray, dash-dotted) for $r_s$ = 0.02, 0.5, 1.0 and 7.0 (blue to red).}
        \end{figure}

        The final model is a linear combination of the left and right regions in Eq.\ (\ref{eqn:im_model}). The left region is weighted by the Fermi function $w = 1 / \{ 1 + \exp[15 \big( k - 1.1 \lambda(r_s, t) \big) ] \}$ and the right side by $1-w$.
        In addition, to prevent spurious negative values at very low $t$, we clip any negative values to be zero. The resulting parametrization is shown in Fig.\ \ref{fig:imag_fit}.
        The absolute mean error for our parametrization is $\approx$ 0.006 $E_F$.
    
\newpage
    \begin{thebibliography}{59}%
      \makeatletter
      \providecommand \@ifxundefined [1]{%
      \@ifx{#1\undefined}
      }%
      \providecommand \@ifnum [1]{%
      \ifnum #1\expandafter \@firstoftwo
      \else \expandafter \@secondoftwo
      \fi
      }%
      \providecommand \@ifx [1]{%
      \ifx #1\expandafter \@firstoftwo
      \else \expandafter \@secondoftwo
      \fi
      }%
      \providecommand \natexlab [1]{#1}%
      \providecommand \enquote  [1]{``#1''}%
      \providecommand \bibnamefont  [1]{#1}%
      \providecommand \bibfnamefont [1]{#1}%
      \providecommand \citenamefont [1]{#1}%
      \providecommand \href@noop [0]{\@secondoftwo}%
      \providecommand \href [0]{\begingroup \@sanitize@url \@href}%
      \providecommand \@href[1]{\@@startlink{#1}\@@href}%
      \providecommand \@@href[1]{\endgroup#1\@@endlink}%
      \providecommand \@sanitize@url [0]{\catcode `\\12\catcode `\$12\catcode
        `\&12\catcode `\#12\catcode `\^12\catcode `\_12\catcode `\%12\relax}%
      \providecommand \@@startlink[1]{}%
      \providecommand \@@endlink[0]{}%
      \providecommand \url  [0]{\begingroup\@sanitize@url \@url }%
      \providecommand \@url [1]{\endgroup\@href {#1}{\urlprefix }}%
      \providecommand \urlprefix  [0]{URL }%
      \providecommand \Eprint [0]{\href }%
      \providecommand \doibase [0]{https://doi.org/}%
      \providecommand \selectlanguage [0]{\@gobble}%
      \providecommand \bibinfo  [0]{\@secondoftwo}%
      \providecommand \bibfield  [0]{\@secondoftwo}%
      \providecommand \translation [1]{[#1]}%
      \providecommand \BibitemOpen [0]{}%
      \providecommand \bibitemStop [0]{}%
      \providecommand \bibitemNoStop [0]{.\EOS\space}%
      \providecommand \EOS [0]{\spacefactor3000\relax}%
      \providecommand \BibitemShut  [1]{\csname bibitem#1\endcsname}%
      \let\auto@bib@innerbib\@empty
      \bibitem [{\citenamefont {Peyrusse}(2008)}]{PEYRUSSE2008}%
        \BibitemOpen
        \bibfield  {author} {\bibinfo {author} {\bibfnamefont {O.}~\bibnamefont
        {Peyrusse}},\ }\href {https://doi.org/10.1088/0953-8984/20/19/195211}
        {\bibfield  {journal} {\bibinfo  {journal} {Journal of Physics: Condensed
        Matter}\ }\textbf {\bibinfo {volume} {20}},\ \bibinfo {pages} {195211}
        (\bibinfo {year} {2008})}\BibitemShut {NoStop}%
      \bibitem [{\citenamefont {Recoules}\ and\ \citenamefont
        {Mazevet}(2009)}]{RECOULES2009}%
        \BibitemOpen
        \bibfield  {author} {\bibinfo {author} {\bibfnamefont {V.}~\bibnamefont
        {Recoules}}\ and\ \bibinfo {author} {\bibfnamefont {S.}~\bibnamefont
        {Mazevet}},\ }\href {https://doi.org/10.1103/PhysRevB.80.064110} {\bibfield
        {journal} {\bibinfo  {journal} {Phys. Rev. B}\ }\textbf {\bibinfo {volume}
        {80}},\ \bibinfo {pages} {064110} (\bibinfo {year} {2009})}\BibitemShut
        {NoStop}%
      \bibitem [{\citenamefont {Cho}\ \emph {et~al.}(2011)\citenamefont {Cho},
        \citenamefont {Engelhorn}, \citenamefont {Correa}, \citenamefont {Ogitsu},
        \citenamefont {Weber}, \citenamefont {Lee}, \citenamefont {Feng},
        \citenamefont {Ni}, \citenamefont {Ping}, \citenamefont {Nelson},
        \citenamefont {Prendergast}, \citenamefont {Lee}, \citenamefont {Falcone},\
        and\ \citenamefont {Heimann}}]{CHO2011}%
        \BibitemOpen
        \bibfield  {author} {\bibinfo {author} {\bibfnamefont {B.~I.}\ \bibnamefont
        {Cho}}, \bibinfo {author} {\bibfnamefont {K.}~\bibnamefont {Engelhorn}},
        \bibinfo {author} {\bibfnamefont {A.~A.}\ \bibnamefont {Correa}}, \bibinfo
        {author} {\bibfnamefont {T.}~\bibnamefont {Ogitsu}}, \bibinfo {author}
        {\bibfnamefont {C.~P.}\ \bibnamefont {Weber}}, \bibinfo {author}
        {\bibfnamefont {H.~J.}\ \bibnamefont {Lee}}, \bibinfo {author} {\bibfnamefont
        {J.}~\bibnamefont {Feng}}, \bibinfo {author} {\bibfnamefont {P.~A.}\
        \bibnamefont {Ni}}, \bibinfo {author} {\bibfnamefont {Y.}~\bibnamefont
        {Ping}}, \bibinfo {author} {\bibfnamefont {A.~J.}\ \bibnamefont {Nelson}},
        \bibinfo {author} {\bibfnamefont {D.}~\bibnamefont {Prendergast}}, \bibinfo
        {author} {\bibfnamefont {R.~W.}\ \bibnamefont {Lee}}, \bibinfo {author}
        {\bibfnamefont {R.~W.}\ \bibnamefont {Falcone}},\ and\ \bibinfo {author}
        {\bibfnamefont {P.~A.}\ \bibnamefont {Heimann}},\ }\href
        {https://doi.org/10.1103/PhysRevLett.106.167601} {\bibfield  {journal}
        {\bibinfo  {journal} {Phys. Rev. Lett.}\ }\textbf {\bibinfo {volume} {106}},\
        \bibinfo {pages} {167601} (\bibinfo {year} {2011})}\BibitemShut {NoStop}%
      \bibitem [{\citenamefont {Dorchies}\ \emph {et~al.}(2015)\citenamefont
        {Dorchies}, \citenamefont {Recoules}, \citenamefont {Bouchet}, \citenamefont
        {Fourment}, \citenamefont {Leguay}, \citenamefont {Cho}, \citenamefont
        {Engelhorn}, \citenamefont {Nakatsutsumi}, \citenamefont {Ozkan},
        \citenamefont {Tschentscher}, \citenamefont {Harmand}, \citenamefont
        {Toleikis}, \citenamefont {St\"ormer}, \citenamefont {Galtier}, \citenamefont
        {Lee}, \citenamefont {Nagler}, \citenamefont {Heimann},\ and\ \citenamefont
        {Gaudin}}]{DORCHIES2015}%
        \BibitemOpen
        \bibfield  {author} {\bibinfo {author} {\bibfnamefont {F.}~\bibnamefont
        {Dorchies}}, \bibinfo {author} {\bibfnamefont {V.}~\bibnamefont {Recoules}},
        \bibinfo {author} {\bibfnamefont {J.}~\bibnamefont {Bouchet}}, \bibinfo
        {author} {\bibfnamefont {C.}~\bibnamefont {Fourment}}, \bibinfo {author}
        {\bibfnamefont {P.~M.}\ \bibnamefont {Leguay}}, \bibinfo {author}
        {\bibfnamefont {B.~I.}\ \bibnamefont {Cho}}, \bibinfo {author} {\bibfnamefont
        {K.}~\bibnamefont {Engelhorn}}, \bibinfo {author} {\bibfnamefont
        {M.}~\bibnamefont {Nakatsutsumi}}, \bibinfo {author} {\bibfnamefont
        {C.}~\bibnamefont {Ozkan}}, \bibinfo {author} {\bibfnamefont
        {T.}~\bibnamefont {Tschentscher}}, \bibinfo {author} {\bibfnamefont
        {M.}~\bibnamefont {Harmand}}, \bibinfo {author} {\bibfnamefont
        {S.}~\bibnamefont {Toleikis}}, \bibinfo {author} {\bibfnamefont
        {M.}~\bibnamefont {St\"ormer}}, \bibinfo {author} {\bibfnamefont
        {E.}~\bibnamefont {Galtier}}, \bibinfo {author} {\bibfnamefont {H.~J.}\
        \bibnamefont {Lee}}, \bibinfo {author} {\bibfnamefont {B.}~\bibnamefont
        {Nagler}}, \bibinfo {author} {\bibfnamefont {P.~A.}\ \bibnamefont
        {Heimann}},\ and\ \bibinfo {author} {\bibfnamefont {J.}~\bibnamefont
        {Gaudin}},\ }\href {https://doi.org/10.1103/PhysRevB.92.144201} {\bibfield
        {journal} {\bibinfo  {journal} {Phys. Rev. B}\ }\textbf {\bibinfo {volume}
        {92}},\ \bibinfo {pages} {144201} (\bibinfo {year} {2015})}\BibitemShut
        {NoStop}%
      \bibitem [{\citenamefont {Engelhorn}\ \emph {et~al.}(2015)\citenamefont
        {Engelhorn}, \citenamefont {Recoules}, \citenamefont {Cho}, \citenamefont
        {Barbrel}, \citenamefont {Mazevet}, \citenamefont {Krol}, \citenamefont
        {Falcone},\ and\ \citenamefont {Heimann}}]{ENGELHORN2015}%
        \BibitemOpen
        \bibfield  {author} {\bibinfo {author} {\bibfnamefont {K.}~\bibnamefont
        {Engelhorn}}, \bibinfo {author} {\bibfnamefont {V.}~\bibnamefont {Recoules}},
        \bibinfo {author} {\bibfnamefont {B.~I.}\ \bibnamefont {Cho}}, \bibinfo
        {author} {\bibfnamefont {B.}~\bibnamefont {Barbrel}}, \bibinfo {author}
        {\bibfnamefont {S.}~\bibnamefont {Mazevet}}, \bibinfo {author} {\bibfnamefont
        {D.~M.}\ \bibnamefont {Krol}}, \bibinfo {author} {\bibfnamefont {R.~W.}\
        \bibnamefont {Falcone}},\ and\ \bibinfo {author} {\bibfnamefont {P.~A.}\
        \bibnamefont {Heimann}},\ }\href {https://doi.org/10.1103/PhysRevB.91.214305}
        {\bibfield  {journal} {\bibinfo  {journal} {Phys. Rev. B}\ }\textbf {\bibinfo
        {volume} {91}},\ \bibinfo {pages} {214305} (\bibinfo {year}
        {2015})}\BibitemShut {NoStop}%
      \bibitem [{\citenamefont {Ogitsu}\ \emph {et~al.}(2018)\citenamefont {Ogitsu},
        \citenamefont {Fernandez-Pañella}, \citenamefont {Hamel}, \citenamefont
        {Correa}, \citenamefont {Prendergast}, \citenamefont {Pemmaraju},\ and\
        \citenamefont {Ping}}]{OGITSU2018}%
        \BibitemOpen
        \bibfield  {author} {\bibinfo {author} {\bibfnamefont {T.}~\bibnamefont
        {Ogitsu}}, \bibinfo {author} {\bibfnamefont {A.}~\bibnamefont
        {Fernandez-Pañella}}, \bibinfo {author} {\bibfnamefont {S.}~\bibnamefont
        {Hamel}}, \bibinfo {author} {\bibfnamefont {A.~A.}\ \bibnamefont {Correa}},
        \bibinfo {author} {\bibfnamefont {D.}~\bibnamefont {Prendergast}}, \bibinfo
        {author} {\bibfnamefont {C.~D.}\ \bibnamefont {Pemmaraju}},\ and\ \bibinfo
        {author} {\bibfnamefont {Y.}~\bibnamefont {Ping}},\ }\href
        {https://doi.org/10.1103/PhysRevB.97.214203} {\bibfield  {journal} {\bibinfo
        {journal} {Phys. Rev. B}\ }\textbf {\bibinfo {volume} {97}},\ \bibinfo
        {pages} {214203} (\bibinfo {year} {2018})}\BibitemShut {NoStop}%
      \bibitem [{\citenamefont {Bolis}\ \emph {et~al.}(2019)\citenamefont {Bolis},
        \citenamefont {Hernandez}, \citenamefont {Recoules}, \citenamefont
        {Guarguaglini}, \citenamefont {Dorchies}, \citenamefont {Jourdain},
        \citenamefont {Ravasio}, \citenamefont {Vinci}, \citenamefont {Brambrink},
        \citenamefont {Ozaki}, \citenamefont {Bouchet}, \citenamefont {Remus},
        \citenamefont {Musella}, \citenamefont {Mazevet}, \citenamefont {Hartley},
        \citenamefont {Guyot},\ and\ \citenamefont {Benuzzi-Mounaix}}]{BOLIS2019}%
        \BibitemOpen
        \bibfield  {author} {\bibinfo {author} {\bibfnamefont {R.}~\bibnamefont
        {Bolis}}, \bibinfo {author} {\bibfnamefont {J.-A.}\ \bibnamefont
        {Hernandez}}, \bibinfo {author} {\bibfnamefont {V.}~\bibnamefont {Recoules}},
        \bibinfo {author} {\bibfnamefont {M.}~\bibnamefont {Guarguaglini}}, \bibinfo
        {author} {\bibfnamefont {F.}~\bibnamefont {Dorchies}}, \bibinfo {author}
        {\bibfnamefont {N.}~\bibnamefont {Jourdain}}, \bibinfo {author}
        {\bibfnamefont {A.}~\bibnamefont {Ravasio}}, \bibinfo {author} {\bibfnamefont
        {T.}~\bibnamefont {Vinci}}, \bibinfo {author} {\bibfnamefont
        {E.}~\bibnamefont {Brambrink}}, \bibinfo {author} {\bibfnamefont
        {N.}~\bibnamefont {Ozaki}}, \bibinfo {author} {\bibfnamefont
        {J.}~\bibnamefont {Bouchet}}, \bibinfo {author} {\bibfnamefont
        {F.}~\bibnamefont {Remus}}, \bibinfo {author} {\bibfnamefont
        {R.}~\bibnamefont {Musella}}, \bibinfo {author} {\bibfnamefont
        {S.}~\bibnamefont {Mazevet}}, \bibinfo {author} {\bibfnamefont {N.~J.}\
        \bibnamefont {Hartley}}, \bibinfo {author} {\bibfnamefont {F.}~\bibnamefont
        {Guyot}},\ and\ \bibinfo {author} {\bibfnamefont {A.}~\bibnamefont
        {Benuzzi-Mounaix}},\ }\href {https://doi.org/10.1063/1.5105390} {\bibfield
        {journal} {\bibinfo  {journal} {Physics of Plasmas}\ }\textbf {\bibinfo
        {volume} {26}},\ \bibinfo {pages} {112703} (\bibinfo {year}
        {2019})}\BibitemShut {NoStop}%
      \bibitem [{\citenamefont {Jourdain}\ \emph {et~al.}(2020)\citenamefont
        {Jourdain}, \citenamefont {Recoules}, \citenamefont {Lecherbourg},
        \citenamefont {Renaudin},\ and\ \citenamefont {Dorchies}}]{JOURDAIN2020}%
        \BibitemOpen
        \bibfield  {author} {\bibinfo {author} {\bibfnamefont {N.}~\bibnamefont
        {Jourdain}}, \bibinfo {author} {\bibfnamefont {V.}~\bibnamefont {Recoules}},
        \bibinfo {author} {\bibfnamefont {L.}~\bibnamefont {Lecherbourg}}, \bibinfo
        {author} {\bibfnamefont {P.}~\bibnamefont {Renaudin}},\ and\ \bibinfo
        {author} {\bibfnamefont {F.}~\bibnamefont {Dorchies}},\ }\href
        {https://doi.org/10.1103/PhysRevB.101.125127} {\bibfield  {journal} {\bibinfo
        {journal} {Phys. Rev. B}\ }\textbf {\bibinfo {volume} {101}},\ \bibinfo
        {pages} {125127} (\bibinfo {year} {2020})}\BibitemShut {NoStop}%
      \bibitem [{\citenamefont {Li}\ \emph {et~al.}(2021)\citenamefont {Li},
        \citenamefont {Li}, \citenamefont {Wang}, \citenamefont {Li}, \citenamefont
        {Kang}, \citenamefont {He},\ and\ \citenamefont {Zhang}}]{ZI2021}%
        \BibitemOpen
        \bibfield  {author} {\bibinfo {author} {\bibfnamefont {Z.}~\bibnamefont
        {Li}}, \bibinfo {author} {\bibfnamefont {W.-J.~L.}\ \bibnamefont {Li}},
        \bibinfo {author} {\bibfnamefont {C.}~\bibnamefont {Wang}}, \bibinfo {author}
        {\bibfnamefont {D.}~\bibnamefont {Li}}, \bibinfo {author} {\bibfnamefont
        {W.}~\bibnamefont {Kang}}, \bibinfo {author} {\bibfnamefont {X.-T.}\
        \bibnamefont {He}},\ and\ \bibinfo {author} {\bibfnamefont {P.}~\bibnamefont
        {Zhang}},\ }\href {https://doi.org/10.1088/1674-1056/abdb1b} {\bibfield
        {journal} {\bibinfo  {journal} {Chinese Physics B}\ }\textbf {\bibinfo
        {volume} {30}},\ \bibinfo {eid} {057102} (\bibinfo {year}
        {2021})}\BibitemShut {NoStop}%
      \bibitem [{\citenamefont {Faleev}\ \emph {et~al.}(2006)\citenamefont {Faleev},
        \citenamefont {van Schilfgaarde}, \citenamefont {Kotani}, \citenamefont
        {L\'eonard},\ and\ \citenamefont {Desjarlais}}]{FALEEV2006}%
        \BibitemOpen
        \bibfield  {author} {\bibinfo {author} {\bibfnamefont {S.~V.}\ \bibnamefont
        {Faleev}}, \bibinfo {author} {\bibfnamefont {M.}~\bibnamefont {van
        Schilfgaarde}}, \bibinfo {author} {\bibfnamefont {T.}~\bibnamefont {Kotani}},
        \bibinfo {author} {\bibfnamefont {F.}~\bibnamefont {L\'eonard}},\ and\
        \bibinfo {author} {\bibfnamefont {M.~P.}\ \bibnamefont {Desjarlais}},\ }\href
        {https://doi.org/10.1103/PhysRevB.74.033101} {\bibfield  {journal} {\bibinfo
        {journal} {Phys. Rev. B}\ }\textbf {\bibinfo {volume} {74}},\ \bibinfo
        {pages} {033101} (\bibinfo {year} {2006})}\BibitemShut {NoStop}%
      \bibitem [{\citenamefont {Hollebon}\ \emph {et~al.}(2019)\citenamefont
        {Hollebon}, \citenamefont {Ciricosta}, \citenamefont {Desjarlais},
        \citenamefont {Cacho}, \citenamefont {Spindloe}, \citenamefont {Springate},
        \citenamefont {Turcu}, \citenamefont {Wark},\ and\ \citenamefont
        {Vinko}}]{HOLLEBON2019}%
        \BibitemOpen
        \bibfield  {author} {\bibinfo {author} {\bibfnamefont {P.}~\bibnamefont
        {Hollebon}}, \bibinfo {author} {\bibfnamefont {O.}~\bibnamefont {Ciricosta}},
        \bibinfo {author} {\bibfnamefont {M.~P.}\ \bibnamefont {Desjarlais}},
        \bibinfo {author} {\bibfnamefont {C.}~\bibnamefont {Cacho}}, \bibinfo
        {author} {\bibfnamefont {C.}~\bibnamefont {Spindloe}}, \bibinfo {author}
        {\bibfnamefont {E.}~\bibnamefont {Springate}}, \bibinfo {author}
        {\bibfnamefont {I.~C.~E.}\ \bibnamefont {Turcu}}, \bibinfo {author}
        {\bibfnamefont {J.~S.}\ \bibnamefont {Wark}},\ and\ \bibinfo {author}
        {\bibfnamefont {S.~M.}\ \bibnamefont {Vinko}},\ }\href
        {https://doi.org/10.1103/PhysRevE.100.043207} {\bibfield  {journal} {\bibinfo
        {journal} {Phys. Rev. E}\ }\textbf {\bibinfo {volume} {100}},\ \bibinfo
        {pages} {043207} (\bibinfo {year} {2019})}\BibitemShut {NoStop}%
      \bibitem [{\citenamefont {Cho}\ \emph {et~al.}(2016)\citenamefont {Cho},
        \citenamefont {Ogitsu}, \citenamefont {Engelhorn}, \citenamefont {Correa},
        \citenamefont {Ping}, \citenamefont {Lee}, \citenamefont {Bae}, \citenamefont
        {Prendergast}, \citenamefont {Falcone},\ and\ \citenamefont
        {Heimann}}]{CHO2016}%
        \BibitemOpen
        \bibfield  {author} {\bibinfo {author} {\bibfnamefont {B.~I.}\ \bibnamefont
        {Cho}}, \bibinfo {author} {\bibfnamefont {T.}~\bibnamefont {Ogitsu}},
        \bibinfo {author} {\bibfnamefont {K.}~\bibnamefont {Engelhorn}}, \bibinfo
        {author} {\bibfnamefont {A.~A.}\ \bibnamefont {Correa}}, \bibinfo {author}
        {\bibfnamefont {Y.}~\bibnamefont {Ping}}, \bibinfo {author} {\bibfnamefont
        {J.~W.}\ \bibnamefont {Lee}}, \bibinfo {author} {\bibfnamefont {L.~J.}\
        \bibnamefont {Bae}}, \bibinfo {author} {\bibfnamefont {D.}~\bibnamefont
        {Prendergast}}, \bibinfo {author} {\bibfnamefont {R.~W.}\ \bibnamefont
        {Falcone}},\ and\ \bibinfo {author} {\bibfnamefont {P.~A.}\ \bibnamefont
        {Heimann}},\ }\href {https://doi.org/10.1038/srep18843} {\bibfield  {journal}
        {\bibinfo  {journal} {Scientific Reports}\ }\textbf {\bibinfo {volume} {6}},\
        \bibinfo {pages} {18843} (\bibinfo {year} {2016})}\BibitemShut {NoStop}%
      \bibitem [{\citenamefont {Denoeud}\ \emph {et~al.}(2014)\citenamefont
        {Denoeud}, \citenamefont {Benuzzi-Mounaix}, \citenamefont {Ravasio},
        \citenamefont {Dorchies}, \citenamefont {Leguay}, \citenamefont {Gaudin},
        \citenamefont {Guyot}, \citenamefont {Brambrink}, \citenamefont {Koenig},
        \citenamefont {Le~Pape},\ and\ \citenamefont {Mazevet}}]{DENOEUD2014}%
        \BibitemOpen
        \bibfield  {author} {\bibinfo {author} {\bibfnamefont {A.}~\bibnamefont
        {Denoeud}}, \bibinfo {author} {\bibfnamefont {A.}~\bibnamefont
        {Benuzzi-Mounaix}}, \bibinfo {author} {\bibfnamefont {A.}~\bibnamefont
        {Ravasio}}, \bibinfo {author} {\bibfnamefont {F.}~\bibnamefont {Dorchies}},
        \bibinfo {author} {\bibfnamefont {P.~M.}\ \bibnamefont {Leguay}}, \bibinfo
        {author} {\bibfnamefont {J.}~\bibnamefont {Gaudin}}, \bibinfo {author}
        {\bibfnamefont {F.}~\bibnamefont {Guyot}}, \bibinfo {author} {\bibfnamefont
        {E.}~\bibnamefont {Brambrink}}, \bibinfo {author} {\bibfnamefont
        {M.}~\bibnamefont {Koenig}}, \bibinfo {author} {\bibfnamefont
        {S.}~\bibnamefont {Le~Pape}},\ and\ \bibinfo {author} {\bibfnamefont
        {S.}~\bibnamefont {Mazevet}},\ }\href
        {https://doi.org/10.1103/PhysRevLett.113.116404} {\bibfield  {journal}
        {\bibinfo  {journal} {Phys. Rev. Lett.}\ }\textbf {\bibinfo {volume} {113}},\
        \bibinfo {pages} {116404} (\bibinfo {year} {2014})}\BibitemShut {NoStop}%
      \bibitem [{\citenamefont {Karasiev}\ \emph {et~al.}(2016)\citenamefont
        {Karasiev}, \citenamefont {Calder\'{\i}n},\ and\ \citenamefont
        {Trickey}}]{Karasiev2016}%
        \BibitemOpen
        \bibfield  {author} {\bibinfo {author} {\bibfnamefont {V.~V.}\ \bibnamefont
        {Karasiev}}, \bibinfo {author} {\bibfnamefont {L.}~\bibnamefont
        {Calder\'{\i}n}},\ and\ \bibinfo {author} {\bibfnamefont {S.~B.}\
        \bibnamefont {Trickey}},\ }\href {https://doi.org/10.1103/PhysRevE.93.063207}
        {\bibfield  {journal} {\bibinfo  {journal} {Phys. Rev. E}\ }\textbf {\bibinfo
        {volume} {93}},\ \bibinfo {pages} {063207} (\bibinfo {year}
        {2016})}\BibitemShut {NoStop}%
      \bibitem [{\citenamefont {Karasiev}\ \emph {et~al.}(2018)\citenamefont
        {Karasiev}, \citenamefont {Dufty},\ and\ \citenamefont {Trickey}}]{KDT16}%
        \BibitemOpen
        \bibfield  {author} {\bibinfo {author} {\bibfnamefont {V.~V.}\ \bibnamefont
        {Karasiev}}, \bibinfo {author} {\bibfnamefont {J.~W.}\ \bibnamefont
        {Dufty}},\ and\ \bibinfo {author} {\bibfnamefont {S.}~\bibnamefont
        {Trickey}},\ }\href@noop {} {\bibfield  {journal} {\bibinfo  {journal} {Phys.
        Rev. Lett.}\ }\textbf {\bibinfo {volume} {120}},\ \bibinfo {pages} {076401}
        (\bibinfo {year} {2018})}\BibitemShut {NoStop}%
      \bibitem [{\citenamefont {Ichimaru}\ \emph {et~al.}(1987)\citenamefont
        {Ichimaru}, \citenamefont {Iyetomi},\ and\ \citenamefont {Tanaka}}]{IIT1987}%
        \BibitemOpen
        \bibfield  {author} {\bibinfo {author} {\bibfnamefont {S.}~\bibnamefont
        {Ichimaru}}, \bibinfo {author} {\bibfnamefont {H.}~\bibnamefont {Iyetomi}},\
        and\ \bibinfo {author} {\bibfnamefont {S.}~\bibnamefont {Tanaka}},\ }\href
        {https://doi.org/https://doi.org/10.1016/0370-1573(87)90125-6} {\bibfield
        {journal} {\bibinfo  {journal} {Physics Reports}\ }\textbf {\bibinfo {volume}
        {149}},\ \bibinfo {pages} {91} (\bibinfo {year} {1987})}\BibitemShut
        {NoStop}%
      \bibitem [{\citenamefont {Tanaka}(2017)}]{IIT2017}%
        \BibitemOpen
        \bibfield  {author} {\bibinfo {author} {\bibfnamefont {S.}~\bibnamefont
        {Tanaka}},\ }\href {https://doi.org/https://doi.org/10.1002/ctpp.201600096}
        {\bibfield  {journal} {\bibinfo  {journal} {Contributions to Plasma Physics}\
        }\textbf {\bibinfo {volume} {57}},\ \bibinfo {pages} {126} (\bibinfo {year}
        {2017})}\BibitemShut {NoStop}%
      \bibitem [{\citenamefont {Tanaka}(2016)}]{TANAKA2016}%
        \BibitemOpen
        \bibfield  {author} {\bibinfo {author} {\bibfnamefont {S.}~\bibnamefont
        {Tanaka}},\ }\href {https://doi.org/10.1063/1.4969071} {\bibfield  {journal}
        {\bibinfo  {journal} {The Journal of Chemical Physics}\ }\textbf {\bibinfo
        {volume} {145}},\ \bibinfo {pages} {214104} (\bibinfo {year}
        {2016})}\BibitemShut {NoStop}%
      \bibitem [{\citenamefont {Perrot}\ and\ \citenamefont
        {Dharma-wardana}(2000)}]{PDW2000}%
        \BibitemOpen
        \bibfield  {author} {\bibinfo {author} {\bibfnamefont {F.}~\bibnamefont
        {Perrot}}\ and\ \bibinfo {author} {\bibfnamefont {M.~W.~C.}\ \bibnamefont
        {Dharma-wardana}},\ }\href {https://doi.org/10.1103/PhysRevB.62.16536}
        {\bibfield  {journal} {\bibinfo  {journal} {Phys. Rev. B}\ }\textbf {\bibinfo
        {volume} {62}},\ \bibinfo {pages} {16536} (\bibinfo {year}
        {2000})}\BibitemShut {NoStop}%
      \bibitem [{\citenamefont {Karasiev}\ \emph {et~al.}(2014)\citenamefont
        {Karasiev}, \citenamefont {Sjostrom}, \citenamefont {Dufty},\ and\
        \citenamefont {Trickey}}]{KSDT2014}%
        \BibitemOpen
        \bibfield  {author} {\bibinfo {author} {\bibfnamefont {V.~V.}\ \bibnamefont
        {Karasiev}}, \bibinfo {author} {\bibfnamefont {T.}~\bibnamefont {Sjostrom}},
        \bibinfo {author} {\bibfnamefont {J.}~\bibnamefont {Dufty}},\ and\ \bibinfo
        {author} {\bibfnamefont {S.~B.}\ \bibnamefont {Trickey}},\ }\href
        {https://doi.org/10.1103/PhysRevLett.112.076403} {\bibfield  {journal}
        {\bibinfo  {journal} {Phys. Rev. Lett.}\ }\textbf {\bibinfo {volume} {112}},\
        \bibinfo {pages} {076403} (\bibinfo {year} {2014})}\BibitemShut {NoStop}%
      \bibitem [{\citenamefont {Groth}\ \emph {et~al.}(2017)\citenamefont {Groth},
        \citenamefont {Dornheim}, \citenamefont {Sjostrom}, \citenamefont {Malone},
        \citenamefont {Foulkes},\ and\ \citenamefont {Bonitz}}]{GDTTFB2017}%
        \BibitemOpen
        \bibfield  {author} {\bibinfo {author} {\bibfnamefont {S.}~\bibnamefont
        {Groth}}, \bibinfo {author} {\bibfnamefont {T.}~\bibnamefont {Dornheim}},
        \bibinfo {author} {\bibfnamefont {T.}~\bibnamefont {Sjostrom}}, \bibinfo
        {author} {\bibfnamefont {F.~D.}\ \bibnamefont {Malone}}, \bibinfo {author}
        {\bibfnamefont {W.~M.~C.}\ \bibnamefont {Foulkes}},\ and\ \bibinfo {author}
        {\bibfnamefont {M.}~\bibnamefont {Bonitz}},\ }\href
        {https://doi.org/10.1103/PhysRevLett.119.135001} {\bibfield  {journal}
        {\bibinfo  {journal} {Phys. Rev. Lett.}\ }\textbf {\bibinfo {volume} {119}},\
        \bibinfo {pages} {135001} (\bibinfo {year} {2017})}\BibitemShut {NoStop}%
      \bibitem [{\citenamefont {Bun\ifmmode~\u{a}\else \u{a}\fi{}u}\ and\
        \citenamefont {Calandra}(2013{\natexlab{a}})}]{XSPECTRA2013}%
        \BibitemOpen
        \bibfield  {author} {\bibinfo {author} {\bibfnamefont {O.}~\bibnamefont
        {Bun\ifmmode~\u{a}\else \u{a}\fi{}u}}\ and\ \bibinfo {author} {\bibfnamefont
        {M.}~\bibnamefont {Calandra}},\ }\href
        {https://doi.org/10.1103/PhysRevB.87.205105} {\bibfield  {journal} {\bibinfo
        {journal} {Phys. Rev. B}\ }\textbf {\bibinfo {volume} {87}},\ \bibinfo
        {pages} {205105} (\bibinfo {year} {2013}{\natexlab{a}})}\BibitemShut
        {NoStop}%
      \bibitem [{\citenamefont {Seah}\ and\ \citenamefont
        {Dench}(1979)}]{SEAHDENCH1979}%
        \BibitemOpen
        \bibfield  {author} {\bibinfo {author} {\bibfnamefont {M.~P.}\ \bibnamefont
        {Seah}}\ and\ \bibinfo {author} {\bibfnamefont {W.~A.}\ \bibnamefont
        {Dench}},\ }\href {https://doi.org/https://doi.org/10.1002/sia.740010103}
        {\bibfield  {journal} {\bibinfo  {journal} {Surface and Interface Analysis}\
        }\textbf {\bibinfo {volume} {1}},\ \bibinfo {pages} {2} (\bibinfo {year}
        {1979})}\BibitemShut {NoStop}%
      \bibitem [{\citenamefont {Cushing}\ \emph {et~al.}(2018)\citenamefont
        {Cushing}, \citenamefont {Zürch}, \citenamefont {Kraus}, \citenamefont
        {Carneiro}, \citenamefont {Lee}, \citenamefont {Chang}, \citenamefont
        {Kaplan},\ and\ \citenamefont {Leone}}]{CUSHING2018}%
        \BibitemOpen
        \bibfield  {author} {\bibinfo {author} {\bibfnamefont {S.~K.}\ \bibnamefont
        {Cushing}}, \bibinfo {author} {\bibfnamefont {M.}~\bibnamefont {Zürch}},
        \bibinfo {author} {\bibfnamefont {P.~M.}\ \bibnamefont {Kraus}}, \bibinfo
        {author} {\bibfnamefont {L.~M.}\ \bibnamefont {Carneiro}}, \bibinfo {author}
        {\bibfnamefont {A.}~\bibnamefont {Lee}}, \bibinfo {author} {\bibfnamefont
        {H.-T.}\ \bibnamefont {Chang}}, \bibinfo {author} {\bibfnamefont {C.~J.}\
        \bibnamefont {Kaplan}},\ and\ \bibinfo {author} {\bibfnamefont {S.~R.}\
        \bibnamefont {Leone}},\ }\href {https://doi.org/10.1063/1.5038015} {\bibfield
        {journal} {\bibinfo  {journal} {Structural Dynamics}\ }\textbf {\bibinfo
        {volume} {5}},\ \bibinfo {pages} {054302} (\bibinfo {year}
        {2018})}\BibitemShut {NoStop}%
      \bibitem [{\citenamefont {Rehr}\ and\ \citenamefont {Albers}(2000)}]{REHR2000}%
        \BibitemOpen
        \bibfield  {author} {\bibinfo {author} {\bibfnamefont {J.~J.}\ \bibnamefont
        {Rehr}}\ and\ \bibinfo {author} {\bibfnamefont {R.~C.}\ \bibnamefont
        {Albers}},\ }\href {https://doi.org/10.1103/RevModPhys.72.621} {\bibfield
        {journal} {\bibinfo  {journal} {Rev. Mod. Phys.}\ }\textbf {\bibinfo {volume}
        {72}},\ \bibinfo {pages} {621} (\bibinfo {year} {2000})}\BibitemShut
        {NoStop}%
      \bibitem [{\citenamefont {Korringa}(1947)}]{KORRINGA1947}%
        \BibitemOpen
        \bibfield  {author} {\bibinfo {author} {\bibfnamefont {J.}~\bibnamefont
        {Korringa}},\ }\href
        {https://doi.org/https://doi.org/10.1016/0031-8914(47)90013-X} {\bibfield
        {journal} {\bibinfo  {journal} {Physica}\ }\textbf {\bibinfo {volume} {13}},\
        \bibinfo {pages} {392} (\bibinfo {year} {1947})}\BibitemShut {NoStop}%
      \bibitem [{\citenamefont {Kohn}\ and\ \citenamefont
        {Rostoker}(1954)}]{KOHN1954}%
        \BibitemOpen
        \bibfield  {author} {\bibinfo {author} {\bibfnamefont {W.}~\bibnamefont
        {Kohn}}\ and\ \bibinfo {author} {\bibfnamefont {N.}~\bibnamefont
        {Rostoker}},\ }\href {https://doi.org/10.1103/PhysRev.94.1111} {\bibfield
        {journal} {\bibinfo  {journal} {Phys. Rev.}\ }\textbf {\bibinfo {volume}
        {94}},\ \bibinfo {pages} {1111} (\bibinfo {year} {1954})}\BibitemShut
        {NoStop}%
      \bibitem [{\citenamefont {Dupree}(1961)}]{DUPREE1961}%
        \BibitemOpen
        \bibfield  {author} {\bibinfo {author} {\bibfnamefont {T.~H.}\ \bibnamefont
        {Dupree}},\ }\href
        {https://doi.org/https://doi.org/10.1016/0003-4916(61)90166-X} {\bibfield
        {journal} {\bibinfo  {journal} {Annals of Physics}\ }\textbf {\bibinfo
        {volume} {15}},\ \bibinfo {pages} {63} (\bibinfo {year} {1961})}\BibitemShut
        {NoStop}%
      \bibitem [{\citenamefont {Beeby}\ and\ \citenamefont
        {Edwards}(1967)}]{BEEBY1967}%
        \BibitemOpen
        \bibfield  {author} {\bibinfo {author} {\bibfnamefont {J.~L.}\ \bibnamefont
        {Beeby}}\ and\ \bibinfo {author} {\bibfnamefont {S.~F.}\ \bibnamefont
        {Edwards}},\ }\href {https://doi.org/10.1098/rspa.1967.0230} {\bibfield
        {journal} {\bibinfo  {journal} {Proceedings of the Royal Society of London.
        Series A. Mathematical and Physical Sciences}\ }\textbf {\bibinfo {volume}
        {302}},\ \bibinfo {pages} {113} (\bibinfo {year} {1967})}\BibitemShut
        {NoStop}%
      \bibitem [{\citenamefont {Morgan}(1966)}]{MORGAN1966}%
        \BibitemOpen
        \bibfield  {author} {\bibinfo {author} {\bibfnamefont {G.~J.}\ \bibnamefont
        {Morgan}},\ }\href {https://doi.org/10.1088/0370-1328/89/2/316} {\bibfield
        {journal} {\bibinfo  {journal} {Proceedings of the Physical Society}\
        }\textbf {\bibinfo {volume} {89}},\ \bibinfo {pages} {365} (\bibinfo {year}
        {1966})}\BibitemShut {NoStop}%
      \bibitem [{\citenamefont {Casida}(1995)}]{Casida1995}%
        \BibitemOpen
        \bibfield  {author} {\bibinfo {author} {\bibfnamefont {M.}~\bibnamefont
        {Casida}},\ }\href@noop {} {\bibfield  {journal} {\bibinfo  {journal} {Phys.
        Rev. A}\ }\textbf {\bibinfo {volume} {51}},\ \bibinfo {pages} {2005}
        (\bibinfo {year} {1995})}\BibitemShut {NoStop}%
      \bibitem [{\citenamefont {Benedict}\ \emph {et~al.}(2002)\citenamefont
        {Benedict}, \citenamefont {Spataru},\ and\ \citenamefont
        {Louie}}]{BENEDICT2002}%
        \BibitemOpen
        \bibfield  {author} {\bibinfo {author} {\bibfnamefont {L.~X.}\ \bibnamefont
        {Benedict}}, \bibinfo {author} {\bibfnamefont {C.~D.}\ \bibnamefont
        {Spataru}},\ and\ \bibinfo {author} {\bibfnamefont {S.~G.}\ \bibnamefont
        {Louie}},\ }\href {https://doi.org/10.1103/PhysRevB.66.085116} {\bibfield
        {journal} {\bibinfo  {journal} {Phys. Rev. B}\ }\textbf {\bibinfo {volume}
        {66}},\ \bibinfo {pages} {085116} (\bibinfo {year} {2002})}\BibitemShut
        {NoStop}%
      \bibitem [{\citenamefont {Kas}\ and\ \citenamefont {Rehr}(2017)}]{KAS2017}%
        \BibitemOpen
        \bibfield  {author} {\bibinfo {author} {\bibfnamefont {J.~J.}\ \bibnamefont
        {Kas}}\ and\ \bibinfo {author} {\bibfnamefont {J.~J.}\ \bibnamefont {Rehr}},\
        }\href {https://doi.org/10.1103/PhysRevLett.119.176403} {\bibfield  {journal}
        {\bibinfo  {journal} {Phys. Rev. Lett.}\ }\textbf {\bibinfo {volume} {119}},\
        \bibinfo {pages} {176403} (\bibinfo {year} {2017})}\BibitemShut {NoStop}%
      \bibitem [{\citenamefont {Allen}\ and\ \citenamefont
        {Mitrović}(1983)}]{ALLEN1983}%
        \BibitemOpen
        \bibfield  {author} {\bibinfo {author} {\bibfnamefont {P.~B.}\ \bibnamefont
        {Allen}}\ and\ \bibinfo {author} {\bibfnamefont {B.}~\bibnamefont
        {Mitrović}}\ }(\bibinfo  {publisher} {Academic Press},\ \bibinfo {year}
        {1983})\ pp.\ \bibinfo {pages} {1--92}\BibitemShut {NoStop}%
      \bibitem [{\citenamefont {Hedin}(1965)}]{HEDIN1965}%
        \BibitemOpen
        \bibfield  {author} {\bibinfo {author} {\bibfnamefont {L.}~\bibnamefont
        {Hedin}},\ }\href {https://doi.org/10.1103/PhysRev.139.A796} {\bibfield
        {journal} {\bibinfo  {journal} {Phys. Rev.}\ }\textbf {\bibinfo {volume}
        {139}},\ \bibinfo {pages} {A796} (\bibinfo {year} {1965})}\BibitemShut
        {NoStop}%
      \bibitem [{\citenamefont {Tan}\ \emph {et~al.}(2021)\citenamefont {Tan},
        \citenamefont {Kas},\ and\ \citenamefont {Rehr}}]{TAN2021}%
        \BibitemOpen
        \bibfield  {author} {\bibinfo {author} {\bibfnamefont {T.~S.}\ \bibnamefont
        {Tan}}, \bibinfo {author} {\bibfnamefont {J.~J.}\ \bibnamefont {Kas}},\ and\
        \bibinfo {author} {\bibfnamefont {J.~J.}\ \bibnamefont {Rehr}},\ }\href
        {https://doi.org/10.1103/PhysRevB.104.035144} {\bibfield  {journal} {\bibinfo
        {journal} {Phys. Rev. B}\ }\textbf {\bibinfo {volume} {104}},\ \bibinfo
        {pages} {035144} (\bibinfo {year} {2021})}\BibitemShut {NoStop}%
      \bibitem [{\citenamefont {Martin}\ \emph {et~al.}(2016)\citenamefont {Martin},
        \citenamefont {Reining},\ and\ \citenamefont
        {Ceperley}}]{MartinReiningCeperley2016}%
        \BibitemOpen
        \bibfield  {author} {\bibinfo {author} {\bibfnamefont {R.~M.}\ \bibnamefont
        {Martin}}, \bibinfo {author} {\bibfnamefont {L.}~\bibnamefont {Reining}},\
        and\ \bibinfo {author} {\bibfnamefont {D.~M.}\ \bibnamefont {Ceperley}},\
        }\href {https://doi.org/10.1017/CBO9781139050807} {\emph {\bibinfo {title}
        {Interacting Electrons: Theory and Computational Approaches}}}\ (\bibinfo
        {publisher} {Cambridge University Press},\ \bibinfo {year}
        {2016})\BibitemShut {NoStop}%
      \bibitem [{\citenamefont {Bun\ifmmode~\u{a}\else \u{a}\fi{}u}\ and\
        \citenamefont {Calandra}(2013{\natexlab{b}})}]{OANA2013}%
        \BibitemOpen
        \bibfield  {author} {\bibinfo {author} {\bibfnamefont {O.}~\bibnamefont
        {Bun\ifmmode~\u{a}\else \u{a}\fi{}u}}\ and\ \bibinfo {author} {\bibfnamefont
        {M.}~\bibnamefont {Calandra}},\ }\href
        {https://doi.org/10.1103/PhysRevB.87.205105} {\bibfield  {journal} {\bibinfo
        {journal} {Phys. Rev. B}\ }\textbf {\bibinfo {volume} {87}},\ \bibinfo
        {pages} {205105} (\bibinfo {year} {2013}{\natexlab{b}})}\BibitemShut
        {NoStop}%
      \bibitem [{\citenamefont {Taillefumier}\ \emph {et~al.}(2002)\citenamefont
        {Taillefumier}, \citenamefont {Cabaret}, \citenamefont {Flank},\ and\
        \citenamefont {Mauri}}]{TAILLEFUMIER2002}%
        \BibitemOpen
        \bibfield  {author} {\bibinfo {author} {\bibfnamefont {M.}~\bibnamefont
        {Taillefumier}}, \bibinfo {author} {\bibfnamefont {D.}~\bibnamefont
        {Cabaret}}, \bibinfo {author} {\bibfnamefont {A.-M.}\ \bibnamefont {Flank}},\
        and\ \bibinfo {author} {\bibfnamefont {F.}~\bibnamefont {Mauri}},\ }\href
        {https://doi.org/10.1103/PhysRevB.66.195107} {\bibfield  {journal} {\bibinfo
        {journal} {Phys. Rev. B}\ }\textbf {\bibinfo {volume} {66}},\ \bibinfo
        {pages} {195107} (\bibinfo {year} {2002})}\BibitemShut {NoStop}%
      \bibitem [{\citenamefont {Gougoussis}\ \emph {et~al.}(2009)\citenamefont
        {Gougoussis}, \citenamefont {Calandra}, \citenamefont {Seitsonen},\ and\
        \citenamefont {Mauri}}]{GOUGOUSSIS2009}%
        \BibitemOpen
        \bibfield  {author} {\bibinfo {author} {\bibfnamefont {C.}~\bibnamefont
        {Gougoussis}}, \bibinfo {author} {\bibfnamefont {M.}~\bibnamefont
        {Calandra}}, \bibinfo {author} {\bibfnamefont {A.~P.}\ \bibnamefont
        {Seitsonen}},\ and\ \bibinfo {author} {\bibfnamefont {F.}~\bibnamefont
        {Mauri}},\ }\href {https://doi.org/10.1103/PhysRevB.80.075102} {\bibfield
        {journal} {\bibinfo  {journal} {Phys. Rev. B}\ }\textbf {\bibinfo {volume}
        {80}},\ \bibinfo {pages} {075102} (\bibinfo {year} {2009})}\BibitemShut
        {NoStop}%
      \bibitem [{\citenamefont {Sham}\ and\ \citenamefont {Kohn}(1966)}]{SHAM1966}%
        \BibitemOpen
        \bibfield  {author} {\bibinfo {author} {\bibfnamefont {L.~J.}\ \bibnamefont
        {Sham}}\ and\ \bibinfo {author} {\bibfnamefont {W.}~\bibnamefont {Kohn}},\
        }\href {https://doi.org/10.1103/PhysRev.145.561} {\bibfield  {journal}
        {\bibinfo  {journal} {Phys. Rev.}\ }\textbf {\bibinfo {volume} {145}},\
        \bibinfo {pages} {561} (\bibinfo {year} {1966})}\BibitemShut {NoStop}%
      \bibitem [{\citenamefont {Mustre~de Leon}\ \emph {et~al.}(1991)\citenamefont
        {Mustre~de Leon}, \citenamefont {Rehr}, \citenamefont {Zabinsky},\ and\
        \citenamefont {Albers}}]{MUSTRE1991}%
        \BibitemOpen
        \bibfield  {author} {\bibinfo {author} {\bibfnamefont {J.}~\bibnamefont
        {Mustre~de Leon}}, \bibinfo {author} {\bibfnamefont {J.~J.}\ \bibnamefont
        {Rehr}}, \bibinfo {author} {\bibfnamefont {S.~I.}\ \bibnamefont {Zabinsky}},\
        and\ \bibinfo {author} {\bibfnamefont {R.~C.}\ \bibnamefont {Albers}},\
        }\href {https://doi.org/10.1103/PhysRevB.44.4146} {\bibfield  {journal}
        {\bibinfo  {journal} {Phys. Rev. B}\ }\textbf {\bibinfo {volume} {44}},\
        \bibinfo {pages} {4146} (\bibinfo {year} {1991})}\BibitemShut {NoStop}%
      \bibitem [{\citenamefont {Ankudinov}\ \emph {et~al.}(1998)\citenamefont
        {Ankudinov}, \citenamefont {Ravel}, \citenamefont {Rehr},\ and\ \citenamefont
        {Conradson}}]{ANKUDINOV1998}%
        \BibitemOpen
        \bibfield  {author} {\bibinfo {author} {\bibfnamefont {A.~L.}\ \bibnamefont
        {Ankudinov}}, \bibinfo {author} {\bibfnamefont {B.}~\bibnamefont {Ravel}},
        \bibinfo {author} {\bibfnamefont {J.~J.}\ \bibnamefont {Rehr}},\ and\
        \bibinfo {author} {\bibfnamefont {S.~D.}\ \bibnamefont {Conradson}},\ }\href
        {https://doi.org/10.1103/PhysRevB.58.7565} {\bibfield  {journal} {\bibinfo
        {journal} {Phys. Rev. B}\ }\textbf {\bibinfo {volume} {58}},\ \bibinfo
        {pages} {7565} (\bibinfo {year} {1998})}\BibitemShut {NoStop}%
      \bibitem [{\citenamefont {Mahan}(2000)}]{MAHAN2000}%
        \BibitemOpen
        \bibfield  {author} {\bibinfo {author} {\bibfnamefont {G.}~\bibnamefont
        {Mahan}},\ }\href@noop {} {\emph {\bibinfo {title} {Many-Particle Physics}}}\
        (\bibinfo  {publisher} {Springer},\ \bibinfo {year} {2000})\BibitemShut
        {NoStop}%
      \bibitem [{\citenamefont {Arista}\ and\ \citenamefont
        {Brandt}(1984)}]{ARISTA1984}%
        \BibitemOpen
        \bibfield  {author} {\bibinfo {author} {\bibfnamefont {N.~R.}\ \bibnamefont
        {Arista}}\ and\ \bibinfo {author} {\bibfnamefont {W.}~\bibnamefont
        {Brandt}},\ }\href {https://doi.org/10.1103/PhysRevA.29.1471} {\bibfield
        {journal} {\bibinfo  {journal} {Phys. Rev. A}\ }\textbf {\bibinfo {volume}
        {29}},\ \bibinfo {pages} {1471} (\bibinfo {year} {1984})}\BibitemShut
        {NoStop}%
      \bibitem [{\citenamefont {Kas}\ \emph {et~al.}(2021)\citenamefont {Kas},
        \citenamefont {Vila}, \citenamefont {Pemmaraju}, \citenamefont {Tan},\ and\
        \citenamefont {Rehr}}]{FEFF10}%
        \BibitemOpen
        \bibfield  {author} {\bibinfo {author} {\bibfnamefont {J.~J.}\ \bibnamefont
        {Kas}}, \bibinfo {author} {\bibfnamefont {F.~D.}\ \bibnamefont {Vila}},
        \bibinfo {author} {\bibfnamefont {C.~D.}\ \bibnamefont {Pemmaraju}}, \bibinfo
        {author} {\bibfnamefont {T.~S.}\ \bibnamefont {Tan}},\ and\ \bibinfo {author}
        {\bibfnamefont {J.~J.}\ \bibnamefont {Rehr}},\ }\href
        {https://doi.org/https://doi.org/10.1107/S1600577521008614} {\bibfield
        {journal} {\bibinfo  {journal} {Journal of Synchrotron Radiation}\ }\textbf
        {\bibinfo {volume} {28}},\ \bibinfo {pages} {1801} (\bibinfo {year}
        {2021})}\BibitemShut {NoStop}%
      \bibitem [{\citenamefont {Rehr}\ \emph {et~al.}(2010)\citenamefont {Rehr},
        \citenamefont {Kas}, \citenamefont {Vila}, \citenamefont {Prange},\ and\
        \citenamefont {Jorissen}}]{FEFF9}%
        \BibitemOpen
        \bibfield  {author} {\bibinfo {author} {\bibfnamefont {J.~J.}\ \bibnamefont
        {Rehr}}, \bibinfo {author} {\bibfnamefont {J.~J.}\ \bibnamefont {Kas}},
        \bibinfo {author} {\bibfnamefont {F.~D.}\ \bibnamefont {Vila}}, \bibinfo
        {author} {\bibfnamefont {M.~P.}\ \bibnamefont {Prange}},\ and\ \bibinfo
        {author} {\bibfnamefont {K.}~\bibnamefont {Jorissen}},\ }\href
        {https://doi.org/10.1039/B926434E} {\bibfield  {journal} {\bibinfo  {journal}
        {Phys. Chem. Chem. Phys.}\ }\textbf {\bibinfo {volume} {12}},\ \bibinfo
        {pages} {5503} (\bibinfo {year} {2010})}\BibitemShut {NoStop}%
      \bibitem [{\citenamefont {Karasiev}\ \emph {et~al.}(2019)\citenamefont
        {Karasiev}, \citenamefont {Trickey},\ and\ \citenamefont
        {Dufty}}]{KTD2019PRB}%
        \BibitemOpen
        \bibfield  {author} {\bibinfo {author} {\bibfnamefont {V.~V.}\ \bibnamefont
        {Karasiev}}, \bibinfo {author} {\bibfnamefont {S.~B.}\ \bibnamefont
        {Trickey}},\ and\ \bibinfo {author} {\bibfnamefont {J.~W.}\ \bibnamefont
        {Dufty}},\ }\href {https://doi.org/10.1103/PhysRevB.99.195134} {\bibfield
        {journal} {\bibinfo  {journal} {Phys. Rev. B}\ }\textbf {\bibinfo {volume}
        {99}},\ \bibinfo {pages} {195134} (\bibinfo {year} {2019})}\BibitemShut
        {NoStop}%
      \bibitem [{FLE()}]{FLEUR}%
        \BibitemOpen
        \href@noop {} {\bibinfo {title} {The {FLEUR} code (version max-6.0):
        www.flapw.de}}\BibitemShut {NoStop}%
      \bibitem [{\citenamefont {Weinert}\ \emph {et~al.}(1982)\citenamefont
        {Weinert}, \citenamefont {Wimmer},\ and\ \citenamefont
        {Freeman}}]{FLPAW1982}%
        \BibitemOpen
        \bibfield  {author} {\bibinfo {author} {\bibfnamefont {M.}~\bibnamefont
        {Weinert}}, \bibinfo {author} {\bibfnamefont {E.}~\bibnamefont {Wimmer}},\
        and\ \bibinfo {author} {\bibfnamefont {A.~J.}\ \bibnamefont {Freeman}},\
        }\href {https://doi.org/10.1103/PhysRevB.26.4571} {\bibfield  {journal}
        {\bibinfo  {journal} {Phys. Rev. B}\ }\textbf {\bibinfo {volume} {26}},\
        \bibinfo {pages} {4571} (\bibinfo {year} {1982})}\BibitemShut {NoStop}%
      \bibitem [{\citenamefont {Alekseeva}\ \emph {et~al.}(2018)\citenamefont
        {Alekseeva}, \citenamefont {Michalicek}, \citenamefont {Wortmann},\ and\
        \citenamefont {Bl{\"u}gel}}]{FLEUR2018}%
        \BibitemOpen
        \bibfield  {author} {\bibinfo {author} {\bibfnamefont {U.}~\bibnamefont
        {Alekseeva}}, \bibinfo {author} {\bibfnamefont {G.}~\bibnamefont
        {Michalicek}}, \bibinfo {author} {\bibfnamefont {D.}~\bibnamefont
        {Wortmann}},\ and\ \bibinfo {author} {\bibfnamefont {S.}~\bibnamefont
        {Bl{\"u}gel}},\ }in\ \href@noop {} {\emph {\bibinfo {booktitle} {Euro-Par
        2018: Parallel Processing}}},\ \bibinfo {editor} {edited by\ \bibinfo
        {editor} {\bibfnamefont {M.}~\bibnamefont {Aldinucci}}, \bibinfo {editor}
        {\bibfnamefont {L.}~\bibnamefont {Padovani}},\ and\ \bibinfo {editor}
        {\bibfnamefont {M.}~\bibnamefont {Torquati}}}\ (\bibinfo  {publisher}
        {Springer International Publishing},\ \bibinfo {address} {Cham},\ \bibinfo
        {year} {2018})\ pp.\ \bibinfo {pages} {735--748}\BibitemShut {NoStop}%
      \bibitem [{\citenamefont {Perdew}\ and\ \citenamefont {Zunger}(1981)}]{PZ1981}%
        \BibitemOpen
        \bibfield  {author} {\bibinfo {author} {\bibfnamefont {J.~P.}\ \bibnamefont
        {Perdew}}\ and\ \bibinfo {author} {\bibfnamefont {A.}~\bibnamefont
        {Zunger}},\ }\href {https://doi.org/10.1103/PhysRevB.23.5048} {\bibfield
        {journal} {\bibinfo  {journal} {Phys. Rev. B}\ }\textbf {\bibinfo {volume}
        {23}},\ \bibinfo {pages} {5048} (\bibinfo {year} {1981})}\BibitemShut
        {NoStop}%
      \bibitem [{\citenamefont {Rehr}\ \emph {et~al.}(1994)\citenamefont {Rehr},
        \citenamefont {Booth}, \citenamefont {Bridges},\ and\ \citenamefont
        {Zabinsky}}]{REHR1994}%
        \BibitemOpen
        \bibfield  {author} {\bibinfo {author} {\bibfnamefont {J.~J.}\ \bibnamefont
        {Rehr}}, \bibinfo {author} {\bibfnamefont {C.~H.}\ \bibnamefont {Booth}},
        \bibinfo {author} {\bibfnamefont {F.}~\bibnamefont {Bridges}},\ and\ \bibinfo
        {author} {\bibfnamefont {S.~I.}\ \bibnamefont {Zabinsky}},\ }\href
        {https://doi.org/10.1103/PhysRevB.49.12347} {\bibfield  {journal} {\bibinfo
        {journal} {Phys. Rev. B}\ }\textbf {\bibinfo {volume} {49}},\ \bibinfo
        {pages} {12347} (\bibinfo {year} {1994})}\BibitemShut {NoStop}%
      \bibitem [{\citenamefont {Kittel}(2005)}]{KITTEL2005}%
        \BibitemOpen
        \bibfield  {author} {\bibinfo {author} {\bibfnamefont {C.}~\bibnamefont
        {Kittel}},\ }\href@noop {} {\emph {\bibinfo {title} {Introduction to solid
        state physics}}},\ \bibinfo {edition} {8th}\ ed.\ (\bibinfo  {publisher}
        {Wiley},\ \bibinfo {address} {Hoboken, NJ},\ \bibinfo {year} {2005})\
        p.~\bibinfo {pages} {20}\BibitemShut {NoStop}%
      \bibitem [{\citenamefont {Kiyono}\ \emph {et~al.}(1978)\citenamefont {Kiyono},
        \citenamefont {Chiba}, \citenamefont {Hayasi}, \citenamefont {Kato},\ and\
        \citenamefont {Mochimaru}}]{KIYONO1978}%
        \BibitemOpen
        \bibfield  {author} {\bibinfo {author} {\bibfnamefont {S.}~\bibnamefont
        {Kiyono}}, \bibinfo {author} {\bibfnamefont {S.}~\bibnamefont {Chiba}},
        \bibinfo {author} {\bibfnamefont {Y.}~\bibnamefont {Hayasi}}, \bibinfo
        {author} {\bibfnamefont {S.}~\bibnamefont {Kato}},\ and\ \bibinfo {author}
        {\bibfnamefont {S.}~\bibnamefont {Mochimaru}},\ }\href
        {https://doi.org/10.7567/jjaps.17s2.212} {\bibfield  {journal} {\bibinfo
        {journal} {Japanese Journal of Applied Physics}\ }\textbf {\bibinfo {volume}
        {17}},\ \bibinfo {pages} {212} (\bibinfo {year} {1978})}\BibitemShut
        {NoStop}%
      \bibitem [{\citenamefont {Kas}\ \emph {et~al.}(2007)\citenamefont {Kas},
        \citenamefont {Sorini}, \citenamefont {Prange}, \citenamefont {Cambell},
        \citenamefont {Soininen},\ and\ \citenamefont {Rehr}}]{KAS2007}%
        \BibitemOpen
        \bibfield  {author} {\bibinfo {author} {\bibfnamefont {J.~J.}\ \bibnamefont
        {Kas}}, \bibinfo {author} {\bibfnamefont {A.~P.}\ \bibnamefont {Sorini}},
        \bibinfo {author} {\bibfnamefont {M.~P.}\ \bibnamefont {Prange}}, \bibinfo
        {author} {\bibfnamefont {L.~W.}\ \bibnamefont {Cambell}}, \bibinfo {author}
        {\bibfnamefont {J.~A.}\ \bibnamefont {Soininen}},\ and\ \bibinfo {author}
        {\bibfnamefont {J.~J.}\ \bibnamefont {Rehr}},\ }\href
        {https://doi.org/10.1103/PhysRevB.76.195116} {\bibfield  {journal} {\bibinfo
        {journal} {Phys. Rev. B}\ }\textbf {\bibinfo {volume} {76}},\ \bibinfo
        {pages} {195116} (\bibinfo {year} {2007})}\BibitemShut {NoStop}%
      \bibitem [{\citenamefont {Kas}\ \emph {et~al.}(2009)\citenamefont {Kas},
        \citenamefont {Vinson}, \citenamefont {Trcera}, \citenamefont {Cabaret},
        \citenamefont {Shirley},\ and\ \citenamefont {Rehr}}]{KAS2009}%
        \BibitemOpen
        \bibfield  {author} {\bibinfo {author} {\bibfnamefont {J.~J.}\ \bibnamefont
        {Kas}}, \bibinfo {author} {\bibfnamefont {J.}~\bibnamefont {Vinson}},
        \bibinfo {author} {\bibfnamefont {N.}~\bibnamefont {Trcera}}, \bibinfo
        {author} {\bibfnamefont {D.}~\bibnamefont {Cabaret}}, \bibinfo {author}
        {\bibfnamefont {E.~L.}\ \bibnamefont {Shirley}},\ and\ \bibinfo {author}
        {\bibfnamefont {J.~J.}\ \bibnamefont {Rehr}},\ }\href
        {https://doi.org/10.1088/1742-6596/190/1/012009} {\bibfield  {journal}
        {\bibinfo  {journal} {Journal of physics. Conference series}\ }\textbf
        {\bibinfo {volume} {190}},\ \bibinfo {pages} {10.1088/1742} (\bibinfo {year}
        {2009})}\BibitemShut {NoStop}%
      \bibitem [{\citenamefont {Lundqvist}(1967)}]{LUNDQVIST1967}%
        \BibitemOpen
        \bibfield  {author} {\bibinfo {author} {\bibfnamefont {B.~I.}\ \bibnamefont
        {Lundqvist}},\ }\href {https://doi.org/10.1007/BF02422717} {\bibfield
        {journal} {\bibinfo  {journal} {Physik der kondensierten Materie}\ }\textbf
        {\bibinfo {volume} {6}},\ \bibinfo {pages} {206} (\bibinfo {year}
        {1967})}\BibitemShut {NoStop}%
      \bibitem [{\citenamefont {Hedin}\ and\ \citenamefont
        {Lundqvist}(1971)}]{HEDIN1971}%
        \BibitemOpen
        \bibfield  {author} {\bibinfo {author} {\bibfnamefont {L.}~\bibnamefont
        {Hedin}}\ and\ \bibinfo {author} {\bibfnamefont {B.~I.}\ \bibnamefont
        {Lundqvist}},\ }\href {https://doi.org/10.1088/0022-3719/4/14/022} {\bibfield
        {journal} {\bibinfo  {journal} {Journal of Physics C: Solid State Physics}\
        }\textbf {\bibinfo {volume} {4}},\ \bibinfo {pages} {2064} (\bibinfo {year}
        {1971})}\BibitemShut {NoStop}%
    \end{thebibliography}
\end{document}